\documentclass[a4paper,usenatbib]{mnras} 

\usepackage{mathptmx}
\usepackage{xspace}
\usepackage{fleqn}

\usepackage[T1]{fontenc}
\usepackage{ae,aecompl}

\usepackage[draft,textsize=footnotesize]{todonotes}%
\usepackage[normalem]{ulem}%

\usepackage{graphicx}	
\usepackage{amsmath}	
\usepackage{amssymb}	

\usepackage{aasmacros}
\usepackage{units}

\newcommand{\Msun}{\mathrm{M_{\sun}}}

\newcommand{\kms}{$\rmn{km\,s^{-1}}$\xspace} 
\newcommand{\vmax}{$V_{\rmn{max}}$\xspace}
\newcommand{\rmax}{$r_{\rmn{max}}$\xspace}

\def\gsim{ \lower .75ex \hbox{$\sim$} \llap{\raise .27ex \hbox{$>$}} }
\def\lsim{ \lower .75ex \hbox{$\sim$} \llap{\raise .27ex \hbox{$<$}} }

\def\jcap{JCAP}

\title[X-ray signals due to decaying dark matter]{The signal of decaying dark matter with hydrodynamical simulations}
\author[M. R. Lovell et al.]{Mark R. Lovell$^{1,2}$\thanks{E-mail: lovell@hi.is}, David Barnes$^{3}$, Yannick Bah\'e$^{4}$,  Joop Schaye$^{4}$, \newauthor Matthieu Schaller$^{4}$, Tom Theuns$^{2}$, Sownak Bose$^{5}$, Robert A. Crain$^{6}$, \newauthor Claudio dalla Vecchia$^{7,8}$, Carlos S. Frenk$^{2}$, Wojciech Hellwing$^{9}$, Scott T. Kay$^{10}$, \newauthor Aaron D. Ludlow$^{11}$ and Richard G. Bower$^{2}$\\
$^{1}$Center for Astrophysics and Cosmology, Science Institute, University of Iceland, Dunhagi 5, 107 Reykjavik, Iceland \\
$^{2}$Institute for Computational Cosmology, Durham University, South Road, Durham DH1 3LE, UK\\
$^{3}$Department of Physics, Kavli Institute for Astrophysics and Space Research, Massachusetts Institute of Technology, Cambridge, MA 02139, USA\\
$^{4}$Leiden Observatory, Leiden University, PO Box 9513, NL-2300 RA Leiden, the Netherlands \\
$^{5}$Harvard-Smithsonian Center for Astrophysics, 60 Garden St., Cambridge, MA 02138, USA\\
$^{6}$Astrophysics Research Institute, Liverpool John Moores University, 146 Brownlow Hill, Liverpool L3 5RF, UK\\
$^{7}$Instituto de Astrof\'isica de Canarias, C/V\'ia L\'actea s/n, E-38205 La Laguna, Tenerife, Spain\\
$^{8}$Departamento de Astrof\'isica, Universidad de La Laguna, Av. del Astrof\'isico Francisco S\'anchez s/n, E-38206 La Laguna, Tenerife, Spain\\
$^{9}$Center for Theoretical Physics, Polish Academy of Sciences, Al.  Lotnik\'ow 32/46, 02-668 Warsaw, Poland\\
${10}$Jodrell Bank Centre for Astrophysics, School of Physics and Astronomy, The University of Manchester, Manchester M13 9PL, UK\\
$^{11}$ICRAR M468, The University of Western Australia, 35 Stirling Hwy, Crawley, Western Australia, 6009\\
}

  \date{Accepted *** Received ***; in original
    form ***} 

\begin{document}

  \pagerange{\pageref{firstpage}--\pageref{lastpage}} \pubyear{2016}

  \maketitle

  \label{firstpage}

  \begin{abstract}
  
  \noindent Dark matter particles may decay, emitting photons. Drawing on the EAGLE family of hydrodynamic simulations of galaxy formation -- including the APOSTLE and C-EAGLE simulations -- we assess the systematic uncertainties and scatter on the decay flux from different galaxy classes, from Milky Way satellites to galaxy clusters, and compare our results to studies of the 3.55~keV line. We demonstrate that previous detections and non-detections of this line are consistent with a dark matter interpretation. For example, in our simulations the width of the the dark matter decay line for Perseus-analogue galaxy clusters lies in the range 1300-1700~\kms. Therefore, the non-detection of the 3.55~keV line in the centre of the Perseus cluster by the {\it Hitomi} collaboration is consistent with detections by other instruments.
We also consider trends with stellar and halo mass and evaluate the scatter in the expected fluxes arising from the anisotropic halo mass distribution and from object-to-object variations.
We provide specific predictions for observations with {\it XMM-Newton} and with the planned X-ray telescopes {\it XRISM} and {\it ATHENA}. If future detections of unexplained X-ray lines match our predictions, including line widths, we will have strong evidence that we have discovered the dark matter. 
     
  \end{abstract}

  \begin{keywords}
    cosmology: dark matter, galaxies: Local Group
  \end{keywords}

  \section{Introduction}
  \label{intro}
 
 One of the main techniques in the toolbox for identifying dark matter is `indirect detection'. This is the detection of products of the decay or annihilation of dark matter particles in astrophysical observations. The best studied mechanism for indirect detection is the annihilation of dark matter particles into a cascade of lower mass particles, ultimately producing photons that are detectable with gamma-ray observatories. This process occurs for $\sim$GeV and heavier weakly interacting massive particles (WIMPs, see \citealp{Arcadi18,Roszkowski18} for a review.) So far no unambiguous signal has been detected, for review see \citet{Gaskins16} and \citet{Slatyer17}. Given that these dark matter candidates have not been detected in complementary direct detection experiments \mbox{\citep[most recently][]{Akerib17,Aprile18}} or collider searches \citep{Aaboud18,Sirunyan18} it is more important than ever to study the possibilities for detecting dark matter models other than WIMPs. 
 
 An alternative mechanism for the indirect detection of dark matter particles is decay. This has received less attention than annihilation  because generic WIMPs would decay very fast unless a symmetry is introduced that ensures its stability \citep[e.g.][]{Pagels82}. because generic WIMP would decay very fast unless a symmetry is introduced that ensures its stability \citep[see e.g.][]{Bobrovskyi11}; however, these theories received much less attention (see \citealp{DeLopeAmigo09} for a discussion of decay in supersymmetric models). 
 
 There exist alternative theories that predict the dark matter particle to have a mass many orders of magnitude below that of WIMPs. The most notable is the neutrino minimal standard model \citep[$\nu$MSM,][]{Asaka05,Laine08,Boyarsky09a} which, in addition to explaining baryogenesis and the origin of neutrino masses, generates a dark matter candidate in the form of the keV-scale sterile neutrino. This particle has a decay channel into a standard model neutrino and an X-ray photon, which may be detected as a line in X-ray spectra with half the rest mass energy of the sterile neutrino. The detection of such a line has been claimed in X-ray observations of M31 \citep{Boyarsky14a}, the GC \citep{Boyarsky15}, deep field observations with {\it Chandra} \citep{Cappelluti17} and {\it Nustar} \citep{Neronov16}, and clusters of galaxies \citep{Boyarsky14a,Bulbul14,Urban:2014yda,Bulbul16,Franse16}; a complete discussion of the status of the 3.55~keV can be found in \citet{Adhikari17}.      
 
One of the major uncertainties in the interpretation of a dark matter decay line is the mass and structure of the dark matter halo of the target galaxy/cluster. Studies typically derive a projected dark matter density by inferring a halo mass and concentration from abundance matching \mbox{\citep{Anderson14}}, or alternatively from dynamical measurements that, however, are made at radii very different from those of the X-ray observations \mbox{\citep[see][for a review]{Boyarsky10}}. They also assume a spherically symmetric dark matter profile, and do not take into account the effects of baryons as predicted by hydrodynamical simulations of galaxy formation. Additional uncertainty in low-mass galaxies arises from the fact that particles like the sterile neutrino behave as warm dark matter (WDM), which suppresses halo concentrations relative to the cold dark matter (CDM) family of models to which most annihilating dark matter candidates belong \mbox{\citep{Colin08, Lovell12,Bose16a}}. 

In order to conclude robustly that any reported signal does indeed originate from dark matter decay, multiple identifications must be made across a wide range of galaxy types and environments; each detection must be consistent with all other detections and take into account the presence of baryons. The goal of this study is to make a self-consistent prediction for the dark matter decay rates -- that is applicable for most viable, decaying dark matter particle candidates -- for a wide variety of galaxies.  

We address the issue of uncertainty in the dark matter distribution in galaxies by calculating the projected dark matter density of astrophysical targets in hydrodynamical simulations of galaxy formation over a comprehensive range of target galaxies. The basis of our work is the suite of EAGLE simulations \mbox{\citep{Schaye15,Crain15}}. In order to examine the full diversity of galaxies and environments, we also consider two further sets of simulations, the APOSTLE simulations of Local Group volumes \mbox{\citep{Sawala16, Fattahi15}} and the C-EAGLE simulations of galaxy clusters \mbox{\citep{Bahe17,Barnes17}}; all these simulations use the EAGLE code and closely related versions of the EAGLE galaxy formation model. We thus predict the relative dark matter decay signal flux across five orders of magnitude in halo mass \footnote{We define our halo mass using the virial mass, $M_{200}$, which is the mass enclosed within the radius that encloses an overdensity 200 times the critical density of the Universe, itself labelled $r_{200}$.} and six orders of magnitude in stellar mass. We also analyze WDM versions of the APOSTLE simulations to take account of the uncertainty introduced by free-streaming of light dark matter particles, and predict the full width-half maximum (FWHM) of the line in the C-EAGLE haloes as a dark matter versus gas origin discriminant. Note that the (CDM) APOSTLE simulations are the same as were used for the dark matter annihilation signal prediction papers of \mbox{\citet{Schaller16}} and \mbox{\citet{Calore15}}, and also the direct detection paper of \mbox{\citet{Bozorgnia16}}; this paper therefore completes the set of dark matter direct and indirect detection signals using APOSTLE. 

This paper is organised as follows. In Section~\ref{sec:sims} we present a summary of the simulations we use. In Section~\ref{sec:meth} we present our method for calculating the dark matter decay rate from different astrophysical targets. Our results are presented in Section~\ref{sec:res}, with subsections providing an overview of galaxy dark matter decay flux measurements, the properties of Local Group galaxies, the Perseus cluster, and the comparison of clusters at different redshifts. We draw our conclusions in Section~\ref{sec:conc}. 
  
  \section{Simulations}
  \label{sec:sims}
  
 The primary simulations used in this study are those performed for the EAGLE project \citep{Schaye15,Crain15,McAlpine16}. This is a suite of simulations of periodic cosmological volumes with a state-of-the-art galaxy formation model. The code is a highly modified version of the {\sc gadget3} code  \mbox{\citep{Springel05}} with a pressure-entropy formulation of SPH \mbox{\citep{Hopkins13}}. The galaxy formation model includes subgrid prescriptions for radiative cooling \mbox{\citep{Wiersma09a}}, stellar evolution \citep{Wiersma09b}, star formation \mbox{\citep{Schaye08}}, black hole formation and mergers \mbox{\citep{Springel05b,RosasGuevara15}}, stellar mass loss, and feedback from star formation and AGN \citep{Booth09,DallaVecchia12}. Dark matter haloes are identified using the friends-of-friends (FoF) algorithm \citep{Davis85} and halo substructure is identified using the {\sc subfind} code \mbox{\citep{Springel01,Dolag09}}. The bound galaxy identified with the largest substructure in each FoF halo is considered as the central galaxy, and the remainder of the galaxies as satellites. Many of our simulations also come with an N-body/DMO counterpart simulation in which all matter is treated as collisionless dark matter. The cosmological parameters are consistent with the \citet{PlanckCP13} values: Hubble parameter $h=H_{0}/(100~\rmn{kms^{-1})}=0.6777$, matter density $\Omega_\rmn{M}=0.307$, dark energy density $\Omega_{\Lambda}=0.693$ and baryon energy density parameter $\Omega_\rmn{b}=0.04825$.  
 
Three varieties of the EAGLE model are used in this study: Reference (Ref), Recalibrated (Recal) and AGNdT9. We outline the reasons for adopting the three different models below; please see Section~2 of \mbox{\citet{Schaye15}} for a comprehensive discussion of the difference between Ref and Recal, and \mbox{Table~\ref{SimTab}} for which simulations use which model. The galaxy formation models used in all simulations, including those used in this paper, cannot be derived from first principles. For example, such an idealised approach would require that we simulate simultaneously the flow of gas around galaxies on very large scales (tens of Mpc) down to the formation of individual stars deep within giant molecular clouds ($\sim$pc), which is not currently computationally feasible. Therefore, these simulations approximate the formation of stars and other small-scale processes using a `subgrid' model while simulating just the large-scale flow of material numerically. The form of the subgrid model cannot always be modelled from first principles, and the efficiency of feedback in particular must be `calibrated' against a series of observations, which in the case of EAGLE are the $z=0.1$ galaxy stellar mass function and the sizes of disc galaxies.  

The calibration is, in practice, at its most accurate for a particular simulation resolution, and therefore we are left with a choice when we want to change the resolution: either to recalibrate the model for the new resolution, which is computationally expensive, or to use the previous calibration and accept a worse fit to the calibration observations. The EAGLE cosmological volumes adopt the first option, namely to have one model for its standard resolution, known as Ref, which was run in a 100~Mpc cube box plus several smaller volumes with the same mass resolution, and a second for its smaller, higher resolution simulation (25~Mpc cube, 8 times better mass resolution) called Recal, or Rec. We use both of these in our work, labelled Ref-L100N1504 and Rec-L25N752 respectively. A third cube (50~Mpc, same mass resolution as the 100~Mpc cube) was run with parameters that were further optimised to improve the hot gas content of the highest mass galaxies. The model derived for this simulation is called AGNdT9, and was used for the C-EAGLE simulations; we also use the (50~Mpc) box from EAGLE in which the model is implemented (AGNdT9-L50N752) in order to constrain systematic differences introduced by this parameter change.        
 
For our study of Local Group analogues we use the APOSTLE project simulations \mbox{\citep{Fattahi15,Sawala16}}. These are 12 zoom-in, hydrodynamical simulations of Local Group analogues using the same code and galaxy formation model as Ref-L100N1504, but with mass resolutions $12\times$ and $144\times$ better than Ref-L100N1504 for the intermediate/medium resolution (AP-MR) and high-resolution (AP-HR) versions of APOSTLE respectively. We also use a version of one APOSTLE volume in which the dark matter is warm rather than cold: low mass $(M_{200}\lsim10^{10}\Msun)$ warm dark matter haloes are less concentrated than CDM haloes of the same mass, and we use these simulations to estimate to what degree the lower central densities suppress the dark matter decay flux. This is a previously unpublished simulation that was performed for one of the volumes at the AP-HR resolution and assumes the most extreme sterile neutrino dark matter model in agreement with the 3.55~keV line (AP-HR-LA11, sterile neutrino mass $M=7$~keV, lepton asymmetry $L_6=11.2$) plus its CDM counterpart (AP-HR-CDM). The AP-HR-LA11 run also comes with a medium resolution version, AP-MR-LA11. For all of these APOSTLE runs the cosmological parameters differ slightly from EAGLE in that they assume the {\it WMAP-7} parameters \citep{wmap11}: Hubble parameter $h=H_{0}/(100~\rmn{kms^{-1}})=0.704$, matter density $\Omega_\rmn{M}=0.272$, dark energy density $\Omega_{\Lambda}=0.728$ and baryon energy density parameter $\Omega_\rmn{b}=0.0455$. 

Much of the observational work on decaying dark matter has involved clusters of galaxies \citep{Boyarsky14a,Bulbul14,HitomiC16}. We therefore also include the 30 C-EAGLE simulations of massive galaxy clusters \mbox{\citep{Bahe17,Barnes17}}. These are also zooms; they were selected to be isolated objects at $z=0$, and were run with the AGNdT9 model. They use the same cosmological parameters as the EAGLE simulations. Finally, many of these simulations were run with DMO counterparts, in which the same initial conditions were used but all of the matter is treated as collisionless dark matter. A brief summary of the properties of all the simulations used here is presented in Table~\ref{SimTab}.

\begin{table*}
    \centering
    \caption{Table of basic simulation properties, from left to right: simulation name, number of simulation volumes, simulation dark matter particle mass $m_\rmn{DM}$, maximum physical softening length $\epsilon$, dark matter model, galaxy formation model, simulation box size (or zoom) and whether we use a DMO counterpart in this study. APOSTLE particle masses vary between volumes and are therefore approximate.}
  	\begin{tabular}{|l|r|c|c|c|c|r|c|}
    	\hline
        Name &\# volumes & $m_\rmn{DM}$~$[\Msun]$ & $\epsilon$ [kpc]& DM model & Galaxy formation model & Box size & DMO version \\
 		\hline
        Ref-L100N1504 & 1 & $9.70\times10^{6}$ & 0.7 & CDM & Ref & 100~Mpc & Y \\
         AGNdT9-L50N752 & 1 & $9.70\times10^{6}$ & 0.7 & CDM & AGNdT9 & 50~Mpc & N \\
       Rec-L25N752 & 1 & $1.21\times10^{6}$  & 0.35 & CDM & Rec & 25~Mpc & Y \\
        \hline
        AP-MR-CDM & 12 & $6\times10^{5}$ & 0.35 & CDM & Ref & Zooms & Y \\
              AP-MR-LA11 & 1 & $6\times10^{5}$ & 0.35  &  $M=7$~keV, $L_{6}=11.2$ & Ref & " & N \\
            AP-HR-CDM & 1 & $5\times10^{4}$ & 0.13 &CDM & Ref & " & N \\
          AP-HR-LA11 & 1 & $5\times10^{4}$ & 0.13 &  $M=7$~keV, $L_{6}=11.2$ &Ref & " & N \\
          \hline
          C-EAGLE &  30 & $9.70\times10^{6}$ & 0.7 & CDM & AGNdT9 & " & Y \\
        \hline
 \end{tabular}

 	\label{SimTab}
 \end{table*}
  
  \section{Mock observations}
  \label{sec:meth}
  
Our goal is to make mock observations of the dark matter distribution of each target. The method we use is very similar to that introduced by \mbox{\citet{Lovell15}}. We present a summary here.
  
To begin, we place a virtual observer at a set distance from the centre of potential of the target cluster / galaxy -- hereafter `the target' -- as calculated by {\sc subfind}. The vector between the target and the observer and the assumed field of view (FoV) over which we take data together define a cone. We determine which of the simulation's dark matter particles are located in the cone, and assume that each dark matter particle is radiating decay photons isotropically at a constant rate. The flux measured by the observer is then the sum of the flux from all dark matter particles within the FoV. In the case of DMO simulations we use all high-resolution particles but subtract the universal baryonic mass fraction before calculating the flux, i.e. dark matter mass $m_\rmn{DM}=(1-\Omega_\rmn{b}/\Omega_\rmn{M})m_\rmn{DMO}$, where $m_\rmn{DMO}$ is the DMO simulation particle mass. If there are $N$ dark matter simulation particles in the FoV, the flux, $F$, is:

\begin{equation}
    F = 1.18\times10^{20}\sum_{i=0}^{N} \frac{m_{\rmn{DM},i}}{M_\rmn{DM}\tau}\frac{1}{4\pi d_{i}^2}~\rmn{counts~s^{-1}cm^{-2}}
    \label{eqn:f1}
\end{equation}

\noindent where $d_{i}$ is the distance between the $i$-th particle and the observer in kpc, $M_\rmn{DM}$ is the mass of the dark matter candidate particle in keV, $\tau$ is the particle lifetime in seconds and $m_{\rmn{DM},i}$ is the mass of the $i$-th simulation dark matter particle in $\Msun$; note that in each of our simulations the high resolution dark matter particles have the same mass so $m_{\rmn{DM},i}\equiv m_{\rmn{DM}}$.   

In almost all of our observations, for both zoom simulations and cosmological volumes, we only consider particles within a spherical aperture of 2~Mpc around the centre of the target, either as the centre of the halo or at some point offset from it. This radius is chosen to be big enough to enclose the virial radii of all our host haloes, and we include all particles within the aperture in our calculations regardless of their halo/subhalo membership. We do not therefore include any contribution from haloes along the line of sight more than 2~Mpc from the target, although we do include additional flux from some neighbouring haloes that overlap with the FoV. We discuss the line-of-sight contribution briefly at the end of Section~\ref{SoS}. The one exception to this rule is our virtual observations of ($z\ge0.1$) clusters, where we instead adopt an aperture of 10~Mpc (see Section.~\ref{subsec:hrc}). In the case of zoom simulations we do not use the low resolution, boundary particles in our calculations.
  
  We consider one current and two upcoming X-ray observatories for our analysis: {\it XMM-Newton}, {\it XRISM} and {\it ATHENA}. For our purposes, we assume that the only difference between these three observatories is the size of the FoV. These are $28'\times28'$, which we approximate as a $28'$ diameter circle, for {\it XMM-Newton} and 3' diameter for {\it XRISM} (compared to a $3'\times3'$ square for the previous {\it Hitomi} mission). The {\it ATHENA} observatory has two instruments with their own FoV: WFI ($40'\times40'$) and X-IFU ($5.3'$ diameter). For most of our results we assume the {\it XMM-Newton} FoV, as the one currently operating observatory, and add results from the {\it XRISM} or either of the {\it ATHENA} instruments for the reasons stated below. To measure the FWHM of the line in Perseus we use the {\it XRISM} FoV since this observatory has a velocity resolution of $<600$~\kms  for {\it XRISM/Resolve} compared to  1500~\kms  for {\it XMM-Newton/RGS}. The {\it ATHENA/XIFU} instrument, launched $>7$~years after {\it XRISM} will have a resolution of 200~\kms  over a slightly larger FoV, whereas the {\it ATHENA/WFI} instrument has a much lower spectral resolution ($\sim10,000$~\kms). We therefore use {\it ATHENA/XIFU} for M31 satellite galaxies where its FoV matches well their characteristic sizes ($\sim500$~pc), and use {\it ATHENA/WFI} for the MW satellites.   
  
  Finally, we introduce our definition of the flux units. The flux is typically measured in counts/s/cm$^2$, and the expected flux depends inversely on the particle mass, $M_\rmn{DM}$ and decay time $\tau$ (equation~\ref{eqn:f1}). The most compelling signal to date for decaying dark matter is the 3.55~keV line, which implies a dark matter particle with a mass of 7.1~keV and a lifetime of $\sim10^{28}$~s. We therefore normalise all of our fluxes to what we would expect in counts/s/cm$^2$ for one of these particles, and refer to this normalisation in the text as:
  
  \begin{equation}
  F_{3.55\rmn{keV}}=1~(7.1\rmn{~keV}/M_\rmn{DM})(10^{28}\rmn{s}/\tau)~\rmn{counts~s^{-1}cm^{-2}}.
\end{equation}  
  
  \section{Results}
  \label{sec:res}
  
  This section is split into discussions of four relevant classes of target for X-ray observations: central galaxies at varying distances, Local Group galaxies, the Perseus cluster, and clusters at higher redshifts ($z\leq0.25$). 
  
  \subsection{Overview: central galaxies}
  \label{subsec:overv}
  
  We begin with an overview of the flux measured for all central galaxies in our simulations, and consider the sources of scatter.
  
  \subsubsection{The decay flux--stellar mass relation and systematic uncertainties}
  
\begin{figure*}
  	\includegraphics[scale=0.60]{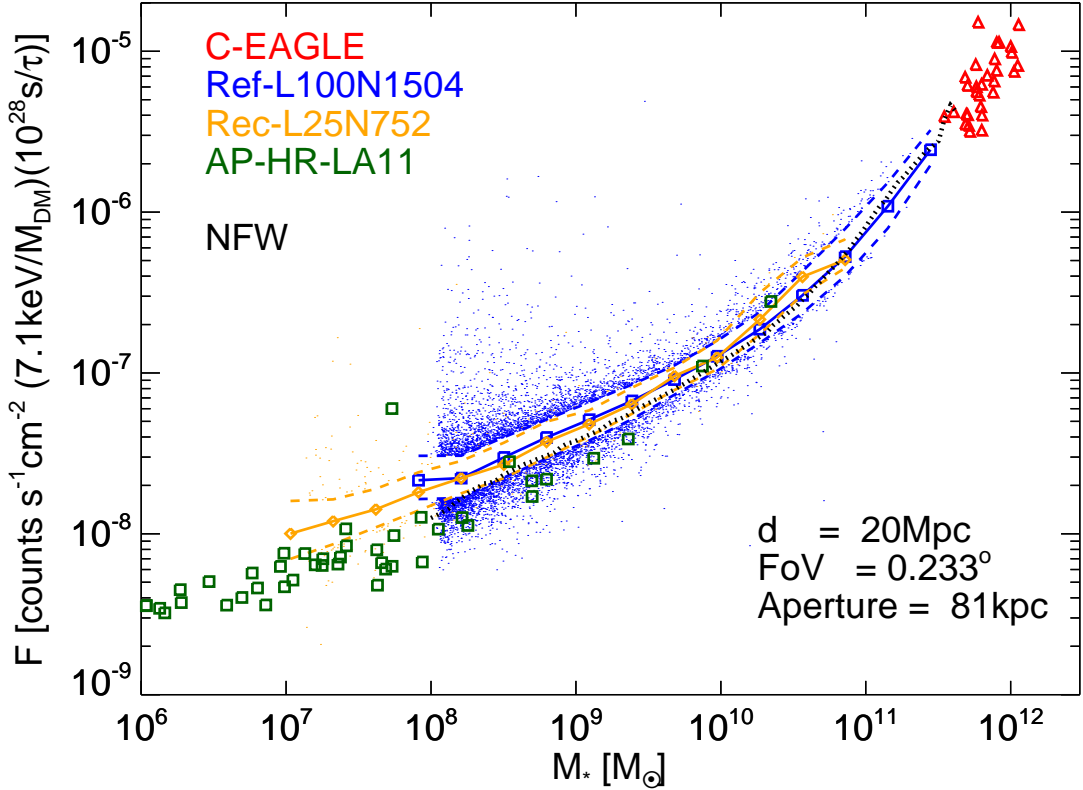}
   \caption{Decay flux as a function of stellar mass for isolated galaxies in EAGLE, APOSTLE and C-EAGLE. We calculate the flux from three orthogonal directions and select the median flux (out of three) for each galaxy. The data sets included are C-EAGLE (red triangles), Ref-L100N1504 (blue), Rec-L25N752 (orange) and AP-HR (green squares). For the two EAGLE volumes, median relations are shown as solid lines, the regions containing 68~per~cent of the data as dashed lines: data points outside these regions are shown as dots. We show the flux-stellar mass relation expected for an NFW profile using the L100N1504-Ref stellar mass-halo mass and halo mass-concentration relations as a dotted black line.}
    \label{MvX}
  \end{figure*}

We first present a common scale of how dark matter decay flux changes with stellar mass for all central galaxies, from $M_{*}=10^{6}\Msun$ dwarf spheroidal galaxies (dSphs) to $M_{*}=10^{12}\Msun$ brightest cluster galaxies (BCGs). In practice, the distances at which galaxies can be observed by flux-limited observations depends strongly on the stellar mass, with dSphs observed no further than 1~Mpc from the Milky Way whereas clusters up to $z=0.35$ (1~Gpc) have been studied in dark matter decay work \mbox{\citep{Bulbul14}}. For our first measurement we therefore place all of our targets at a single distance that is intermediate between the regime of dSphs and that of clusters; we select a proper distance of 20~Mpc, which corresponds to a radius at the target of $\sim80$~kpc for the {\it XMM-Newton} FoV. We draw our targets from the $z=0$ output snapshots of Ref-L100N1504, Rec-L25N752, C-EAGLE and AP-HR-LA11 ($L_6$=11.2); see Table~\ref{SimTab}. We perform three observations of each isolated galaxy in three orthogonal directions. Here, `isolated' galaxies are defined as being the most massive galaxy within their parent FoF halo and also having no other more massive galaxies whose centre-of-potential is within the FoV. We select the median from each set of three flux measurements and plot the results in Fig.~\ref{MvX}, together with a semi-analytic estimate for the flux described below.

  The data sets form a continuous band from a flux of $5\times10^{-9}$~$F_\rmn{3.55keV}$ at $M_{*}=10^{6}\Msun$ to $10^{-5}$~$F_\rmn{3.55keV}$ for the $M_{*}=10^{12}\Msun$ galaxies. At the low mass end of the Ref-L100N1504 dataset there is a considerable upturn in the number of galaxies with very high fluxes, often over ten times the median flux. This effect is at least in part due to nearby massive galaxies that are not centred within the line-of-sight to our target but are nevertheless close enough to contribute additional flux. We have checked this possibility by drawing a spherical aperture with a radius of four virial radii around each galaxy, and removing from our sample any additional galaxies that are located within that aperture: we find that the choice of four virial radii preferentially removes the high flux--low mass galaxies.    
  
  We compare these results to a semi-analytic decay flux-stellar mass relation, first as a simple check of our method and second to show the merits of our particle-based calculations over the semi-analytic approach. We compute the semi-analytic curve as follows. We convolve the median  stellar mass-halo mass relation of the Ref-L100N1504 simulation \mbox{\citep[][ fig.~8b]{Schaye15}} with a power law fit to the halo mass-halo concentration relation of the same simulation \mbox{\citep[][ fig.~11c]{Schaller15}} to obtain the median values of $M_{200}$ and Navarro--Frenk--White profile \mbox{\citep[NFW,][]{NFW_96,NFW_97}} halo concentration, $c$, as a function of stellar mass. Note that the concentration is calculated by fitting NFW profiles to the dark matter components of the \emph{hydrodynamical} Ref-L100N1504 haloes, and therefore accounts for the dark matter halo response to the baryon physics. Having found the pair of $M_{200}-c$ parameters that correspond to each stellar mass, we compute the flux of an NFW profile with that pair of halo parameters for stellar masses in the range $[10^{8},10^{11.3}\Msun]$ and include the result in Fig.~\ref{MvX}. The NFW curve is in good agreement with our simulation results, thus corroborating our direct particle-based method. The agreement is best for the most massive Ref-L100N1504 haloes and progressively underestimates our measured median flux for lower masses, which we expect is due to the presence of neighbouring haloes contributing to the decay flux over and above what the NFW result predicts. We expand on this comparison in Section~\ref{SoS}. 

\begin{figure}
      	\includegraphics[scale=0.33]{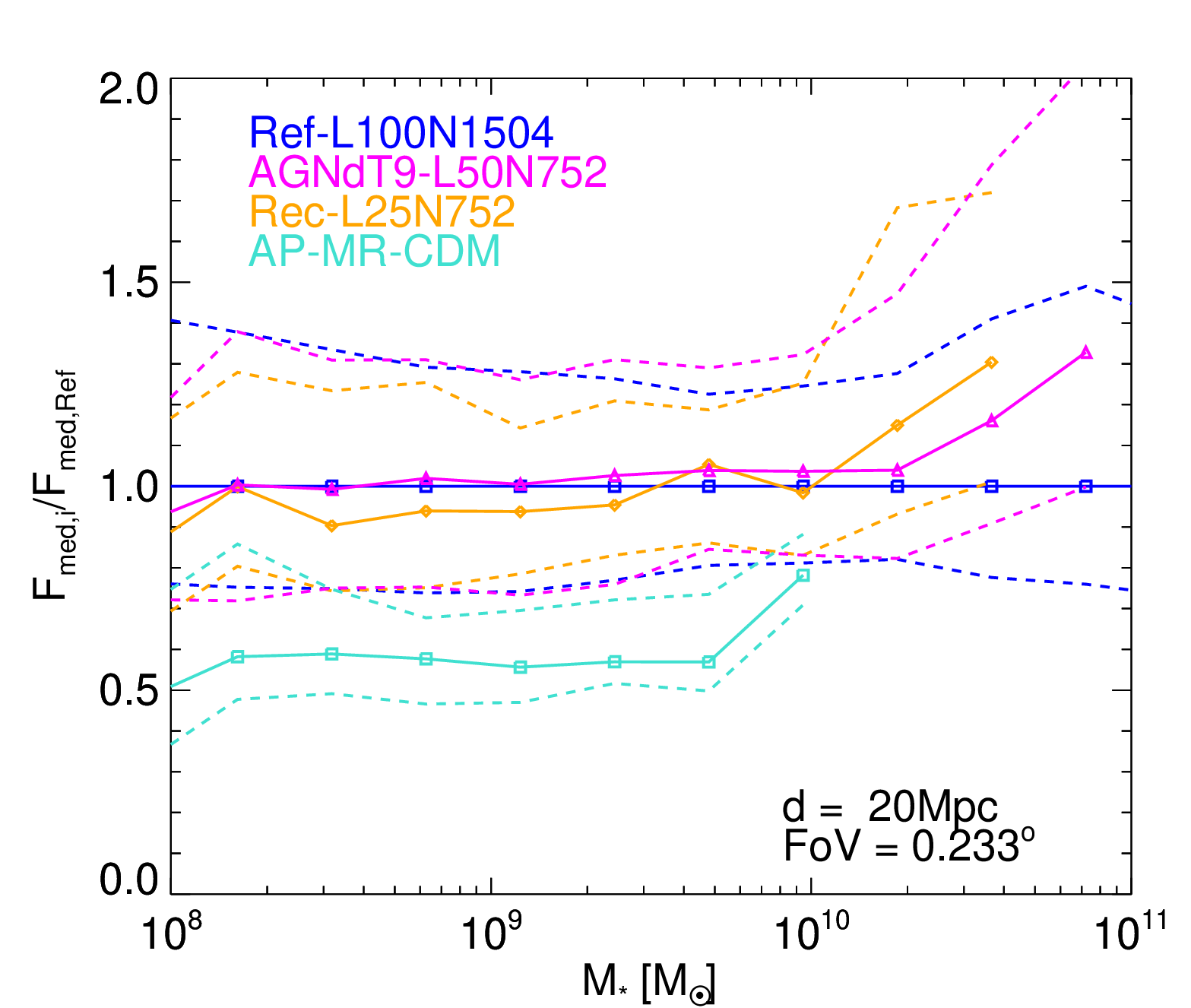}
      \includegraphics[scale=0.33]{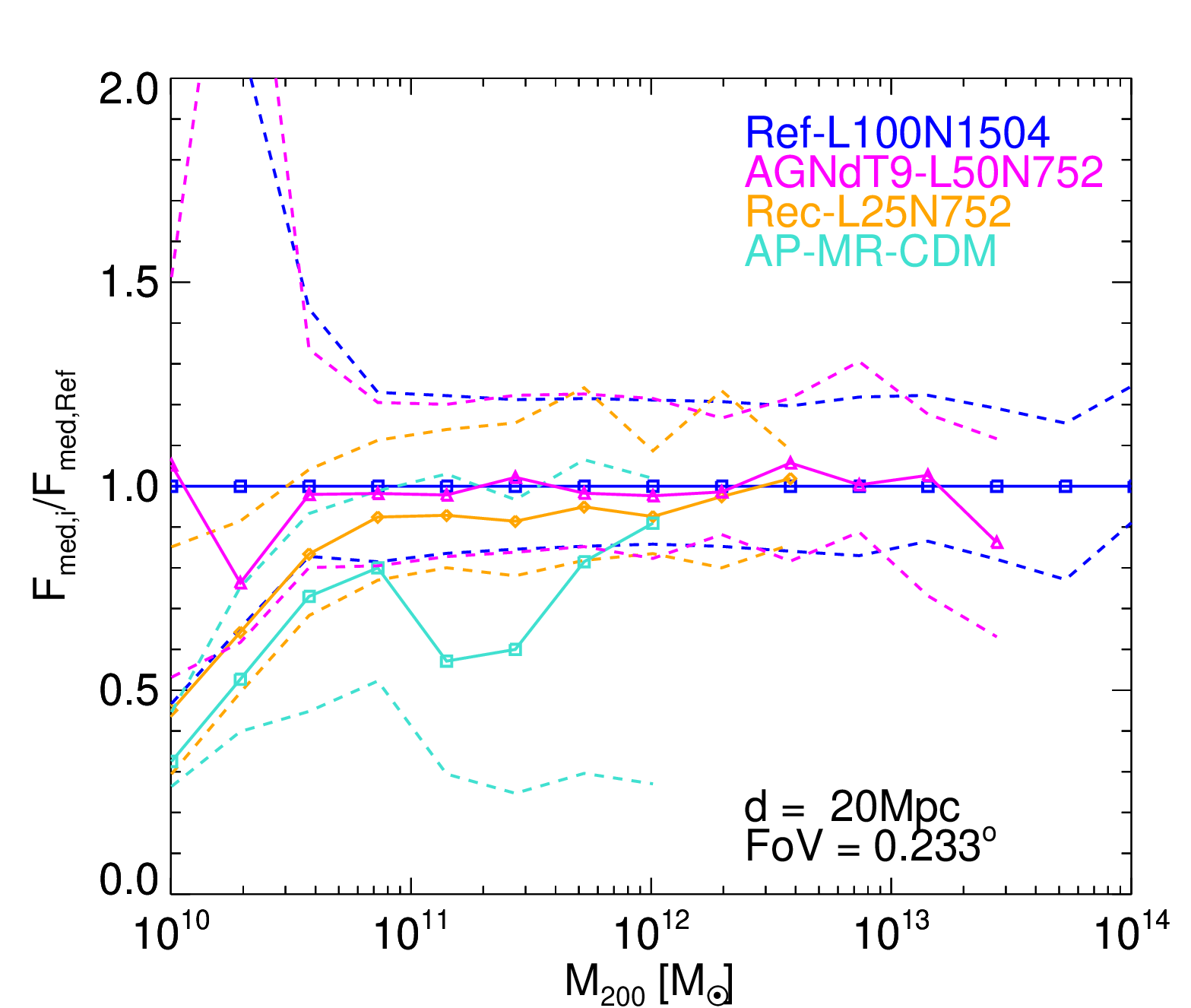}
   \caption{The median decay flux relations of AGNTd9-L50N752 (magenta), Ref-L100N1504 (blue), Rec-L25N752 (orange) and AP-MR-CDM (turquoise)  divided by the median Ref-L100N1504 relation as a function of stellar mass (top panel) and halo mass (bottom panel). The solid lines show the median relations and the dashed lines show the $1\sigma$ scatter.}
    \label{MvXS}
  \end{figure}

  The Ref-L100N1504 and Rec-L25N752 median decay flux--stellar mass relations agree well with each other, but disagree by a factor of two with AP-HR despite the fact AP-HR and Ref-L100N1504 were both run with the Ref model. We explore these differences further, and also make predictions for the expected scatter in flux of these galaxies, in Fig.~\ref{MvXS}, in which we normalise three of our flux relations by that of Ref-L100N1504. We include Rec-L25N752 directly from Fig.~\ref{MvX}, but replace C-EAGLE and AP-HR with two related simulations that contain more galaxies: AGNdT9-L50N752, which was run with the same mass resolution and model parameters as C-EAGLE but in a 50~Mpc periodic volume, and the AP-MR-CDM simulations that use the same galaxy formation model as AP-HR (both CDM and LA11) but with a similar mass resolution to Rec-L25N752. In the same Figure we also show results calculated as a function of halo mass, $M_{200}$, instead of stellar mass.
  
The fluxes predicted by AP-MR-CDM at fixed stellar mass are 40~per~cent lower than those of Ref-L100N1504 compared to less than 10~per~cent lower in Rec-L25N752, which has a similiar resolution to AP-MR-CDM. This is due to the excess stellar mass that is formed at this mass resolution when the Ref galaxy formation model is applied, owing to its lower feedback efficiency \citep{Schaye15}. It follows that at fixed halo mass the stellar mass is higher, and thus at fixed stellar mass the halo mass - and thus total dark matter content -- is lower. Therefore, the difference between AP-MR-CDM and Ref-L100N1504 is smaller when measured at fixed halo mass than at fixed stellar mass case except for a prominent, unexplained dip at $2\times10^{11}\Msun$. 

The AGNdT9-L50N752 simulation shows excellent agreement with Ref-L100N1504 up to $2\times10^{12}\Msun$, above which it diverges to higher fluxes than predicted by up to 30~per~cent at $10^{11}\Msun$ in Ref-L100N1504. This is in spite of the fact that the C-EAGLE haloes show a slightly lower flux per unit stellar mass than one would extrapolate from the bright end of the Ref-L100N1504 in Fig.~\ref{MvX}. The lower flux at fixed stellar mass of C-EAGLE clusters is likely linked to the excessive star formation in BCGs compared to observations \citep{Bahe17} shifting data points to the right. On the other hand, the origin of the excess flux in AGNdT9-L50N752 $M_{*}>2\times10^{10}$ galaxies over their Ref-L100N1504 counterparts is unclear; we speculate that the AGNdT9 model is the more accurate model in this stellar mass range because it produces the better match to the $z=0.1$ stellar mass function \citep[][ fig.~4]{Schaye15}. We conclude that the decay flux measured as a function of stellar mass is affected by the star formation efficiency at the tens of per~cent level for current galaxy formation models, and it is therefore crucial to use an accurately calibrated feedback model when making these predictions.

\begin{figure}
  	\includegraphics[scale=0.33]{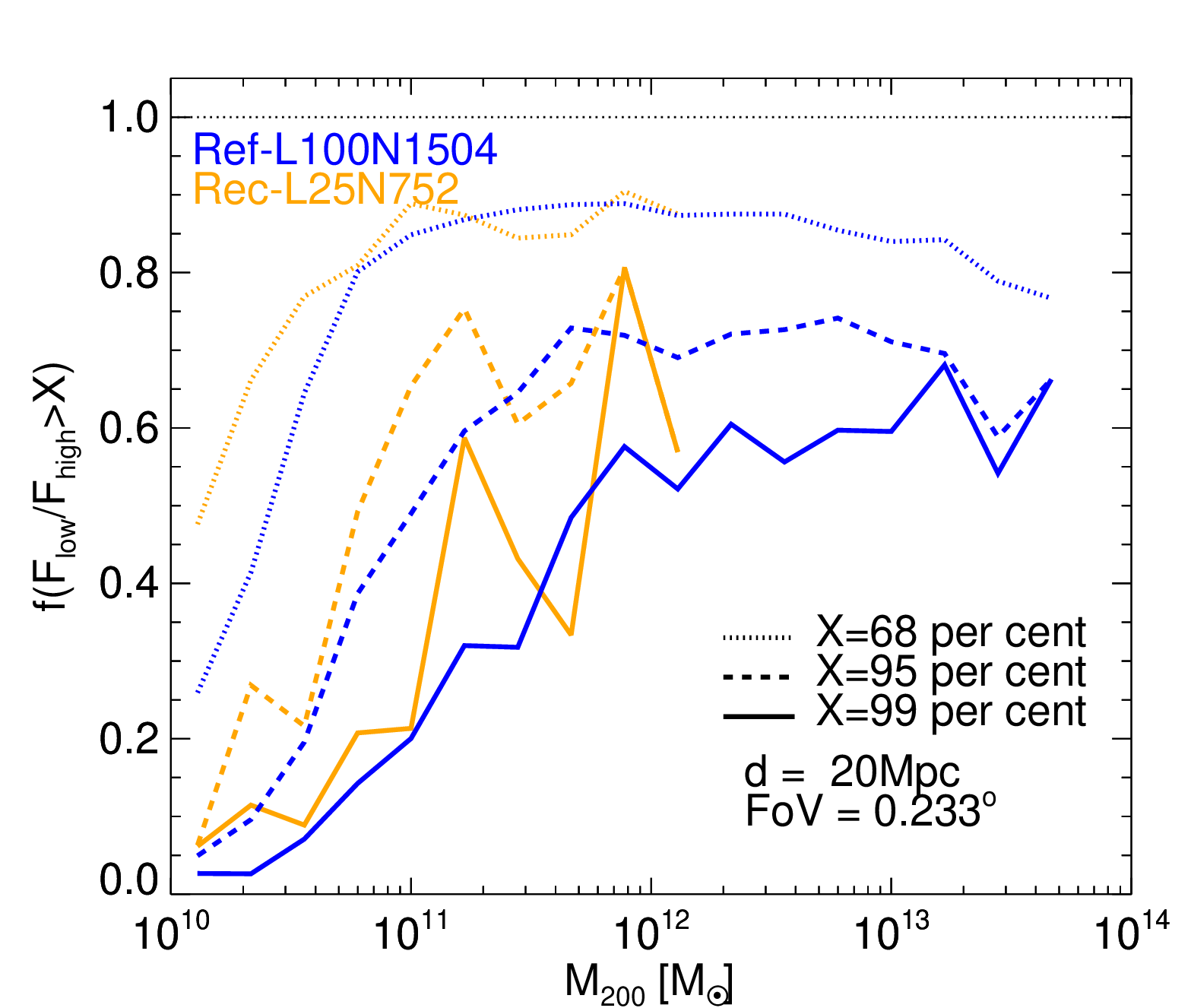}
    \caption{Decay flux ratios of minimum to maximum flux, out of three orthogonal  sightlines for each halo, as a function of halo mass for isolated galaxies. The data sets included are Ref-L100N1504 (blue) and Rec-L25N752 (orange). We calculate the flux from three orthogonal directions and select the lowest and highest flux for each galaxy. The dotted lines show the flux ratio above which 68~per~cent of the data lie, followed by 95~per~cent (dashed lines) and 99~per~cent (solid lines).}
    \label{MvXR}
  \end{figure}

 Fig.~\ref{MvXS} also shows the scatter in the decay flux at fixed stellar mass, which for Ref-L100N1504 is consistently around 30~per~cent ($1\sigma$ scatter). By taking the median flux out of three sightlines, this measurement neglected some portion of the scatter due to the asphericity of the dark matter distribution, which can be caused by different halo shapes, the presence of substructure and local haloes centred outside the FoV that are large enough to contribute mass inside the FoV. We quantify the systematic uncertainty due to this asphericity. We compute the ratio of the lowest to highest flux of the three virtual observations of each galaxy and plot the results in Fig.~\ref{MvXR}, in this case as a function of halo mass rather than stellar mass.

  In general, the variation between directions can be substantial. The smallest variations occur in the most massive haloes ($M_{200}>10^{12}\Msun$), where the difference between the lowest and highest fluxes is $<40$~per~cent for 99~per~cent of galaxies. The variation between orthogonal sightlines increases systematically as halo mass decreases: at $M_{200}=10^{10}\Msun$, we find 70~per~cent suppression in the lowest-to-highest flux ratio at $1\sigma$, 90~per~cent suppression at $2\sigma$ and up to 95~per~cent suppression in the flux between sightlines at 99~per~cent of the data. These results are in good agreement with those reported by \mbox{\citet{Bernal16}}, who performed a similar exercise with the Illustris simulations \mbox{\citep{Vogelsberger14}}. There is remarkably good agreement between the Rec-L25N752 and Ref-L100N1504 simulations at all masses where they both have good statistics except for at $M_{200}<10^{11}\Msun$, where Ref-L100N1504 fluxes show up to 30~per~cent more variation than the Rec-L25N752 galaxies. This indicates the contribution from massive, nearby haloes not present in the small Rec-L25N752 volume as discussed in the context of Fig.~\ref{MvX}.
  
  We have checked for the possibility that the variation of the decay flux with viewing angle is related to the asymmetry of the host halo in the following manner. We computed the dot product of the viewing angle with the minor and major axis vectors of the ellipsoid defined by the inertia tensor of each host halo's dark matter component, obtained the cosine of the subtending angle associated with that dot product, and looked for correlation with measured flux. We found no such correlation between the angle cosine and decay flux, both when using major/minor axis vectors associated with the smooth SUBFIND halo and the larger friends-of-friends halo that contains substructures; we therefore do not find any evidence that the scatter is due to halo triaxiality. We consider an alternative source of scatter, that of satellite galaxies, in Section~\ref{SoS}.

\begin{figure}
  	\includegraphics[scale=0.34]{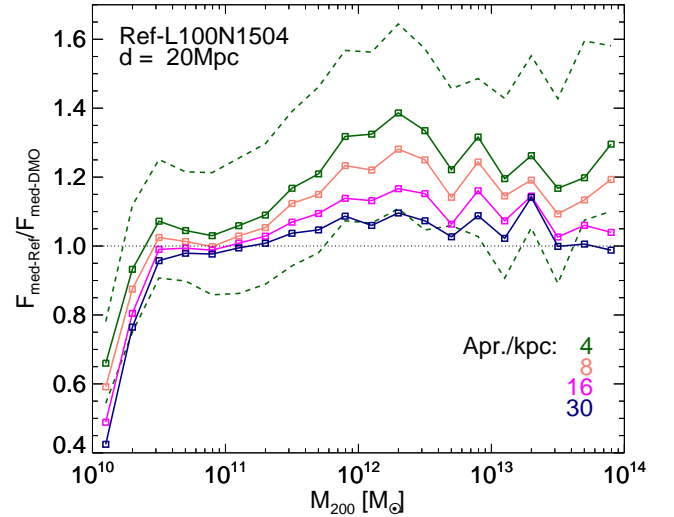}
    \caption{The change in flux due to baryonic effects. We show the median decay flux-halo mass relations for Ref-L100N1504 divided by the medians of their DMO counterparts for four apertures at the target: 4~kpc (green), 8~kpc (pink), 16~kpc (magenta) and 30~kpc (dark blue). The decay flux-halo mass for the DMO simulation is calculated using the DMO-measured decay flux and the baryonic physics counterpart-measured halo mass. The dashed lines show the 68~per~cent scatter on the data (see main text, plotted for 4~kpc only). The targets were placed at a distance of 20~Mpc from the observer.}
    \label{MvXDMO}
  \end{figure}

  The final source of systematic uncertainty on the X-ray decay flux that we consider is the effect of baryons on the dark matter \citep[e.g.][]{Schaller15,Dutton16,Peirani17,Lovell18b}. For example, cooling and subsequent contraction of the gas draws dark matter inward, while repeated, short bursts of star formation can remove enough gas to change the potential and make the dark matter expand outwards \citep{NEF96,Pontzen_Governato_11}. We analyse the effect of baryons on the dark matter by matching haloes between our Ref-L100N1504 run and its DMO counterpart using particle IDs, in order to: i) make sure our halo selections are comparable e.g. with regards to environment, and ii) attach the values of $M_{200}$ for our hydrodynamical haloes to their DMO counterparts in order to eliminate the change in $M_{200}$ due to baryonic physics \citep{Schaller15};  and perform our virtual observations also on the DMO haloes. The net result is two decay flux-halo mass relations, one of which includes baryonic effects on the dark matter distribution and one that does not. In contrast to our previous virtual observations, rather than using the entire FoV of one of the instruments we instead select four aperture radii at the centre of the target -- 4, 8, 16 and 30~kpc -- and compute the flux from these four apertures with an expectation that the effect of baryons is stronger at smaller radii. We place our target galaxies at 20~Mpc from the observer: the 30~kpc aperture then subtends an angle that is approximately the same size as the {\it ATHENA/X-IFU} FoV. Our results are shown in Fig.~\ref{MvXDMO}   
  
  At low halo masses, the DMO counterparts of our $M_{200}<10^{11}\Msun$ Ref-L100N1504 runs have a higher flux for $M_{200}<3\times10^{10}\Msun$, but we anticipate that this result is due to a numerical effect in the hydro run calculation as argued in the context of Fig.~\ref{MvXS}. For larger halo masses than this, the flux in the hydro galaxies increases relative to their DMO counterparts, by up to an average of 40~per~cent enhancement in the 4~kpc aperture at $M_{200}=2\times10^{12}\Msun$. This shows that the measurement of the flux in M31 is likely to be affected by contraction of the halo, an effect that we explore further in Section~\ref{subsec:m31}. The difference between the hydrodynamical and DMO results is systematically smaller with increasing aperture size. We therefore conclude that adiabatic contraction of the dark matter has a measurable impact on the predicted decay flux and therefore makes the decay flux profile steeper than predicted by, for example, the NFW profile.    

  \subsubsection{Sources of scatter}
  \label{SoS}
The origin of the scatter in the mock X-ray flux between galaxies at fixed stellar mass  is important to understand in and of itself, and where that scatter correlates with an observable quantity can be used to further test whether any potential signal is more or less likely to originate from dark matter decay, e.g. in the abundance of bright satellites as shown below. We therefore examine the relationship of galaxy and host halo properties with the X-ray decay flux in Ref-L100N1504 galaxies; we have checked that, in general, the same results are obtained in each case with the Rec-L25N752 simulation, and comment on differences as and when they occur. We perform the first part of the analysis using the full {\it XMM-Newton} FoV (80~kpc aperture at 20~Mpc distance) and the second part with an inner 8~kpc aperture at the same 20~Mpc distance. 

\begin{figure*}
  		\includegraphics[scale=0.66]{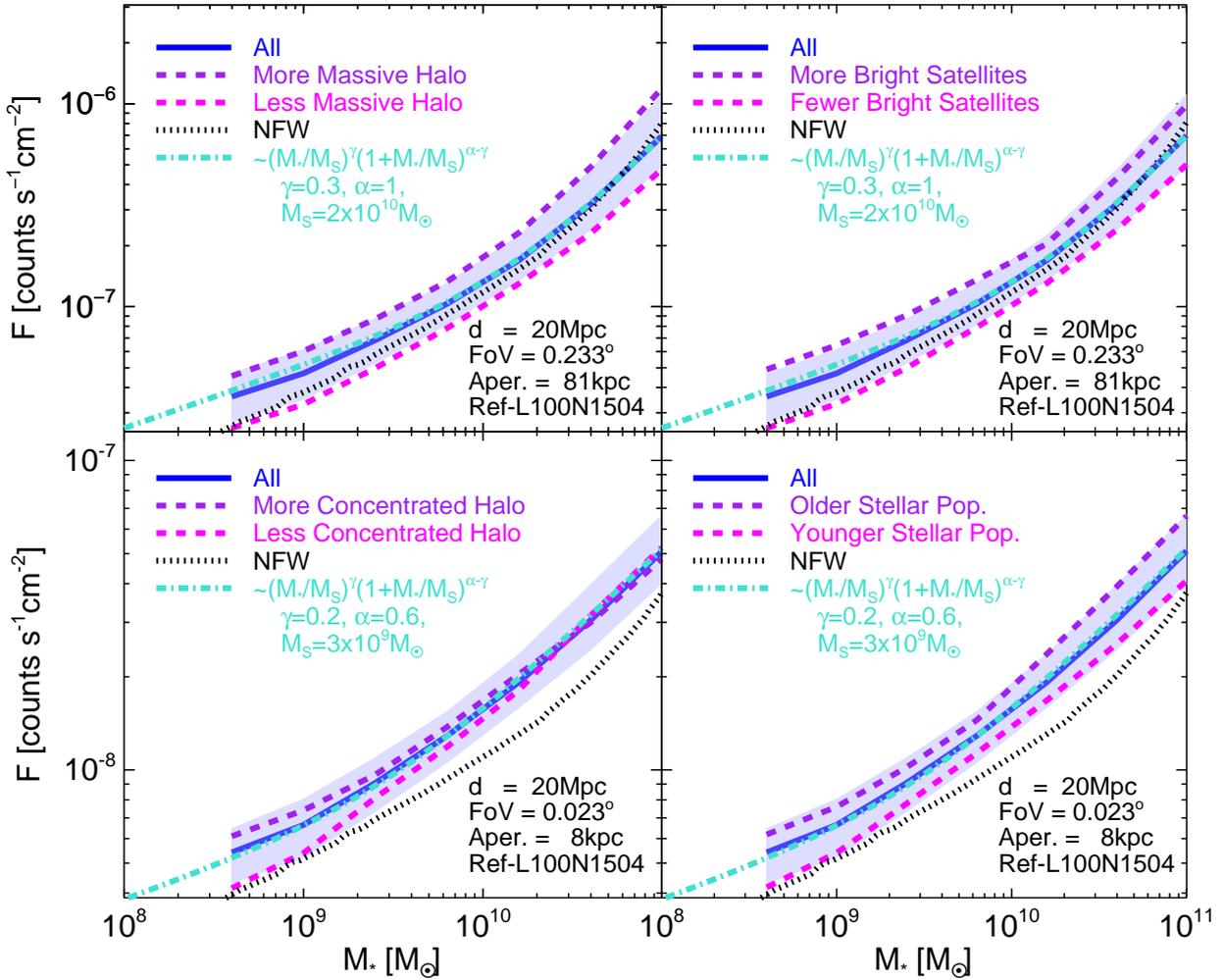}
        \caption{The decay flux of Ref-L100N1504 haloes separated into high and low quartiles in different galaxy/host halo properties (different panels). The population median is shown as a solid blue line and 68~per~cent of the data as a shaded blue region. The upper and lower quartiles for each property are shown as the purple and magenta dashed lines, respectively. The galaxy properties for each panel are: $M_{200}$ (top left), number of satellites with stellar mass at least 10~per~cent of that of the host galaxy (top right), halo concentration $\delta_\rmn{V}$ (bottom left) and the median stellar population age (bottom right). The fluxes are calculated at an observer distance of 20~Mpc; the top two panels use the full {\it XMM-Newton} FoV for an aperture of 81~kpc, and the bottom panels a smaller aperture of 8~kpc. The NFW expectation based on the Ref-L100N1504 stellar mass-halo mass relation and the halo mass-concentration relation described in connection to Fig.~\ref{MvX} is shown as a dotted black line. A double power law fit to the data is shown as a dot-dashed turquoise line, and its equation is given in the Figure legends.}
        \label{Sep4Panel}
  \end{figure*}

We consider four quantities of interest for our galaxies: the host halo mass, $M_{200}$, the number of bright satellites (defined below), the host halo concentration as parametrized by $\delta_{V}=2(V_\rmn{max}/(H_{0}r_\rmn{max}))^{2}$, where \vmax is the peak of the halo circular velocity curve, and \rmax is the radius at which that peak occurs, and the median age of the stellar population; we also allude to other quantities as appropriate. All are presented in Fig.~\ref{Sep4Panel}.

We begin by computing the median decay flux, calculated at 20~Mpc, of Ref-L100N1504 galaxies as a function of stellar mass; we choose 20~Mpc since it is roughly half way between the nearest and most distant galaxies in the \mbox{\citet{Anderson14}} sample and the aperture, 81~kpc, probes much of the physical extent of the host halo. We bin the galaxies by stellar mass, and in each bin calculate the median flux of those galaxies in the upper and lower quartiles of halo mass, $M_{200}$. We present the results in the top left panel of Fig.~\ref{Sep4Panel}, along with the NFW expected stellar mass-flux relation derived for Fig.~\ref{MvX}. We also include an analytic fit to the data as a turquoise line, which we describe below.  
  
  The upper quartile in $M_{200}$ tracks the upper edge of the 68~per~cent region of the galaxy population (shaded region), and in the same manner the lower $M_{200}$ quartile tracks the bottom of the 68~per~cent region. The same pattern occurs when the flux is measured at distances of 10 and 2~Mpc (not shown), and also for the $V_\rmn{max}$ parametrisation of halo mass. We therefore confirm that the scatter in $M_{*}/M_{200}$ is responsible for much of the scatter in the flux at fixed stellar mass.
  
  The halo mass is difficult to measure directly for individual galaxies, and we therefore consider a proxy for this quantity to aid future comparisons with observations. We choose as our proxy the number of bright satellite galaxies, which we define as those bound satellites of the central galaxy (identified by {\sc subfind}) that have a stellar mass of at least 10~per~cent of the central galaxy's stellar mass. We repeat the quartile split performed above for $M_{200}$ using the number of bright satellites, and show the results in the top right panel of Fig.~\ref{Sep4Panel}. The high- and low satellite number subsamples reproduce almost exactly the $M_{200}$ results, as expected from the tight halo mass-substructure abundance relation. We therefore have a means to check any proposed dark matter decay origin using satellite counts, whilst cautioning that observational methods of identifying satellite galaxies are very different to that used by our subhalo finder. 
  
  At this stage we take the opportunity to develop a fitting function for the median flux as a function of stellar mass  assuming Ref-L100N1504 and using the full {\it XMM-Newton} FoV. We obtain a fit for a double power law of the form:
  
  \begin{equation}
  	F = F_{0} (M_{*}/M_\rmn{S})^{\gamma}(1+M_{*}/M_\rmn{S})^{\alpha-\gamma},
    \label{eqn:ff}
	\end{equation}
    \noindent
with power law indicies $\gamma=0.3$, $\alpha=1$, transition mass $M_\rmn{S}=2\times10^{10}\Msun$ and normalisation $F_{0}=1.2\times10^{-7}\rmn{~counts~s^{-1}~cm^{-2}}$. The curve has a slope of index 0.3 for $M_{*}<M_\rmn{S}$ and index 1.0 for $M_{*}>M_\rmn{S}$, and encodes both the halo mass-concentration and stellar mass halo mass relations. We normalise the curve to the measured median value at $M_\rmn{S}$, and obtain agreement between the median and this fit to better than 10~per~cent in the plotted stellar mass range and better than 5~per~cent in the interval $[2.5,100]\times10^{9}\Msun$. This fit also works well above $M_\rmn{S}$ for Rec-L25N752, but overpredicts the fluxes of low mass galaxies in that simulation by up to a factor of 2. We repeat this exercise for the 8~kpc aperture measurements at the end of this Subsection.

  The second fundamental property of a galaxy's host halo, after its mass, is its concentration. Higher concentration haloes will have higher dark matter decay rates when stellar mass and halo mass are fixed simultaneously, as a greater proportion of the dark matter is centrally concentrated and therefore located within the FoV. However, halo mass is anti-correlated with concentration, so in the case that stellar mass alone is fixed, and not halo mass, we expect that more concentrated haloes will exhibit \emph{less} flux than their low-concentration counterparts given the positive correlation of $M_{200}$ with decay flux demonstrated in Fig.~\ref{Sep4Panel}. We check this assertion in the regime where the centre of the halo has the highest contribution relative to its outer parts, namely for the smaller aperture of 8~kpc. We parametrise the concentration using the $\delta_{V}$ parameter and show the results in the bottom left panel of Fig.~\ref{Sep4Panel}.  

Contrary to the simple picture suggested above, we find that for this small aperture low mass ($<1\times10^{10}\Msun$) galaxies exhibit a slight positive correlation between concentration and decay flux that grows stronger to smaller masses. This result likely derives from two sources. The first is the discrepancy between the `true' dark matter profile of simulated dark matter haloes and the model NFW profile in the inner regions of haloes, as was shown for both the EAGLE simulations and their DMO counterparts in \mbox{\citet{Schaller15}}. The difference in the stellar mass-flux relation for the 8~kpc aperture, as shown in the dotted line, is typically 50~per~cent or more for most halo masses, compared to less than 10~per~cent for 81~kpc (c.f. the top two panels of Fig.~\ref{Sep4Panel}). Second, the definition of the concentration scales with the size of the halo whereas the aperture size at the target is fixed. The influence of the concentration of the low mass haloes can therefore be different to that of the high mass haloes. Finally, we have reproduced this experiment for the full 81~kpc aperture and in that case recovered the expected anti-correlation between decay flux and concentration.

We conclude our detailed discussion of secondary quantities with a study of a quantity that is influenced by both halo mass and and concentration, but is more readily observable than either: the median age of the galactic stellar population. Haloes whose inner parts collapse at an earlier time have a higher central density (which is the same as concentration but only at fixed halo mass) and a larger fraction of old stars \citep{Bray16}. We therefore expect galaxies with older stellar populations to exhibit higher dark matter decay fluxes. We define the stellar age of a galaxy as the median age of its constituent star particles, the observational equivalent of which is the median age of its stellar population. We split the Ref-L100N1504 galaxy population -- 8~kpc aperture -- into quartiles based on stellar age in the same manner as for halo mass, satellite counts and concentration, and present our results in the bottom right panel of Fig.~\ref{Sep4Panel}.
  
The galaxies with older stellar populations do indeed exhibit higher decay fluxes, as we argued above, and the correlation is almost as strong as for halo mass. The scatter related to stellar ages is weakest around $M_{*}\sim5\times10^{9}\Msun$, and we have found in the 81~kpc aperture version of this plot (not shown) that the correlation between decay flux and stellar age at this stellar mass disappears completely. However, at the highest and lowest stellar masses the correlation between decay flux and stellar age persists, retaining the values measured at the 8~kpc aperture. 

We note that the fitting function parameters presented in equation~(\ref{eqn:ff}) give a poor fit to our 8~kpc aperture measurements, which is unsurprising given that the outer regions of the halo are not included in this case. We find a better fit is obtained 
with the same formula using $\gamma=0.2$, $\alpha=0.6$, $M_\rmn{S}=3\times10^{9}\Msun$ and $F_{0}=7.5\times10^{-9}\rmn{~counts~s^{-1}~cm^{-2}}$. 

\begin{figure}
	\includegraphics[scale=0.55]{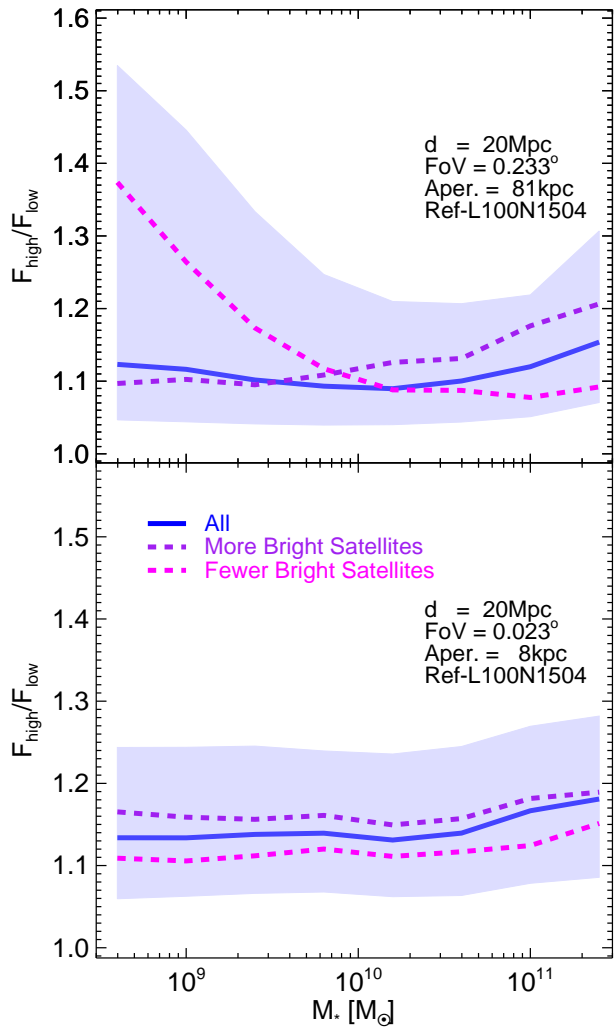}
    \caption{The decay high-to-low flux ratio of Ref-L100N1504 galaxies separated into high and low quartiles by the number of bright satellite galaxies. The fluxes are measured at a distance of 20~Mpc, using the full {\it XMM-Newton} FoV (81~kpc aperture, top panel) and one reduced aperture (8~kpc, bottom panel). The lines and shaded regions indicate the same quantities as in Fig.~\ref{Sep4Panel}, except that fluxes are replaced by flux ratios between viewing angles.}
    \label{MvXFR}
\end{figure}

The bright satellites mentioned above can be expected to correlate with the scatter of the galaxies between viewing angles, as massive satellites will contribute extra dark matter decay flux \mbox{\citep{Bernal16}}. We examine to what degree this is true for our Ref-L100N1504 galaxy sample by measuring the decay flux for three sightlines that are orthogonal to one another per galaxy, computing the ratio of the highest flux to lowest flux, and then repeating the same process as for the brightest satellites panel of Fig.~\ref{Sep4Panel} while replacing the decay flux with the high-to-low flux ratio. We present our results in Fig.~\ref{MvXFR}.   

The median change in flux between our viewing angles for each galaxy is of order 15~per~cent for the 8~kpc aperture measurements and slightly lower, $~\sim12$~per~cent, for the full {\it XMM-Newton} FoV with a potential, weak positive correlation with stellar mass. At $M_{*}>10^{11}\Msun$, there is a preference for galaxies with more satellites to show a greater difference between the two sightlines than those that have fewer, typically by 18~per~cent to 10~per~cent, in qualitative agreement with \mbox{\citet{Bernal16}}. This trend continues consistently to lower stellar masses for the 8~kpc measurements. However, in the 81~kpc case the roles are reversed below $M_{*}=10^{10}\Msun$, with satellite-poor galaxies showing a variation of up to 30~per~cent between sightlines compared to 10~per~cent for satellite-rich systems. We speculate that this fact reflects the change in halo mass relative to nearby haloes: satellite-poor galaxies inhabit less massive haloes, which then receive a higher contribution of flux within one of the three sightlines from neighbouring haloes.  

The final source of scatter that we consider briefly is the presence of dark matter along the line-of-sight that is unassociated with the target, and may contribute to the measured flux. We have estimated the size of this contribution by choosing 500 sightlines that cross the Ref-L100N1504 with a length of 100~Mpc and calculating the measured flux while taking into account the redshifting of the decay flux line due to peculiar velocities and the Hubble expansion. Only a fifth of the sightlines defined encompassed any particles; those that did returned a median flux of $2\times10^{-10} \rm{counts/s/cm^{2}}$, some two orders of magnitude lower than most of our virtual observations and also two orders of magnitude fainter than the decay flux obtained from the uniform critical density of dark matter. We expect that a WDM version of Ref-L100N1504 would show a higher decay background because less of the mass has collapsed into small haloes, but will nevertheless be limited by the uniform critical density, and will therefore not affect our results.        
 
  \subsubsection{Variation in flux with distance}
  \label{sss:dist}
\begin{figure}
  	\includegraphics[scale=0.34]{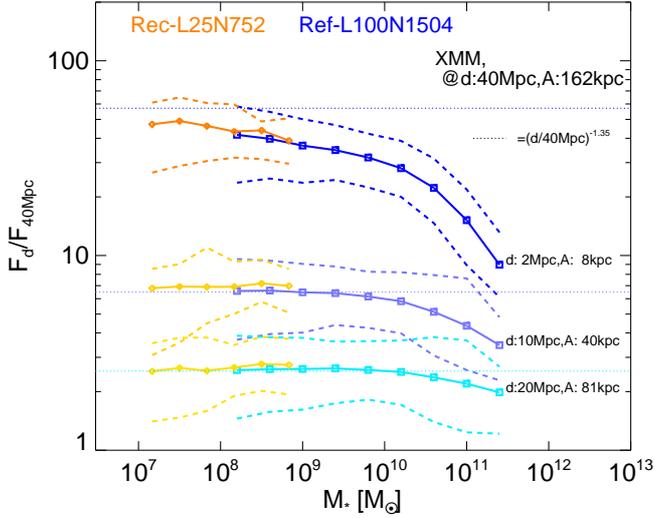}
    \caption{The ratio of the decay flux-stellar mass relation for galaxies observed at 2, 10 and 20~Mpc relative to 40~Mpc using the {it XMM-Newton} FoV. Each ratio is identified by the legend on the right-hand side of the plot. Solid lines show the ratio of the median relations and the dashed lines indicate the 68~per~cent scatter. The Ref-L100N1504 results are shown in blue (2~Mpc), purple (10~Mpc) and cyan (20~Mpc); the Rec-L25N752 as orange, light orange and yellow curves respectively. We limit the stellar mass range of overlap between the two simulations to improve legibility. The radius enclosed by the FoV at each distance is indicated by a letter `A'. We mark the value of the ratio $(d/40~\rmn{Mpc})^{-1.35}$ at each distance with a dotted line.}
    \label{MvXDistR}
  \end{figure}

  We have shown that the dark matter flux for a galaxy with a given stellar mass depends somewhat on intrinsic, correlated factors (halo mass/substructure) and on the implementation of the baryon model (halo mass-stellar mass relation, degree of dark matter contraction). One further factor that is not intrinsic or model dependent, yet is important, is the distance to the target galaxy. The precise distribution of matter within the target, coupled to the size of the instrumental FoV, affects how each galaxy's decay flux declines with distance, at least when the full FoV is considered. We therefore consider four sets of distances as suggested by the X-ray catalogue assembled by \mbox{\citet{Anderson14}}: 2, 10, 20 and 40~Mpc. We place each of our central target galaxies at these four distances and compute the median flux as a function of stellar mass. We then compute the ratio of the 2, 10 and 20~Mpc median relations to that at the largest distance we consider, 40~Mpc, where the size of the aperture subtended by the source plane is larger than the NFW scale radius of most of the haloes considered and thus the results are more easily interpreted. We obtain a 68~per~cent scatter on this relation by taking the ratio of individual 2--10--20~Mpc observations with respect to 40~Mpc observations at the same stellar mass drawn at random (with replacement). We perform this procedure for Ref-L100N1504 and Rec-L25N752, using the {\it XMM-Newton} FoV and plot the results in Fig.~\ref{MvXDistR}.     
  
 In the 10~Mpc and 2~Mpc cases, the ratio of the fluxes drops sharply for stellar masses $>10^{10}\Msun$. At lower stellar masses, the drop off is shallower for the 2~Mpc sample, while the 10 and 20~Mpc trends are almost flat with $M_{*}$. We note that, empirically, the drop off in flux between 10 and 40~Mpc for $M_{*}<10^{10}\Msun$ falls approximately like a power law as $\propto d^{-1.35}$, compared to $\propto d^{-2}$ for a point source. Between 10 and 20~Mpc a still tighter agreement is obtained with $\propto d^{-1.25}$. The transition from a flat relation to one that is falling at higher masses occurs roughly at the peak of star formation efficiency,  $2\times10^{10}\Msun$: towards lower stellar masses than this, the median dark matter host halo is changing mass less rapidly than the stellar mass so the relation is flat, but towards higher masses it is instead the dark halo mass that increases faster per unit log stellar mass \footnote{We have successfully replicated this result using the convolution of the stellar mass-halo mass relation and the mass-concentration relations presented in Fig.~\ref{MvX} and expanded upon in Fig.~\ref{Sep4Panel}}. Recalling equation~(\ref{eqn:ff}), we have therefore shown that the flux for a galaxy of distance $[10,40]$~Mpc and stellar mass $[3,1000]\times10^8\Msun$ measured with the full {\it XMM-Newton} FoV is approximately:
 
  \begin{equation}
  \begin{aligned}
  	F =&7.0\times10^{-6} \left(\frac{d}{\rmn{Mpc}}\right)^{-1.35}\left(\frac{M_{*}}{M_\rmn{S}}\right)^{0.3}\left(1+\frac{M_{*}}{M_\rmn{S}}\right)^{0.7}\times \\ 
    &\left(\frac{7.1~\rmn{keV}}{M_\rmn{DM}}\right)\left(\frac{10^{28}~\rmn{s}}{\tau}\right)\rmn{counts~s}^{-1}\rmn{cm}^{-2},
   \end{aligned}
   \label{eqn:ff2}
	\end{equation}
    
    \noindent while repeating that a better fit between [10,20]~Mpc is obtained with $d^{-1.25}$.
 
 We have also repeated this exercise for the {\it XRISM} and {\it ATHENA/XIFU} instruments, which probe different parts of the halo profile due to their smaller FoV and approximate a subregion of the {\it XMM-Newton} FoV. We find the variations with distance when using the {\it XRISM} instrument are quite different to those obtained with {\it XMM-Newton}. The variation with stellar mass is much steeper, and the change in the mean drop off in flux is better described by a power law of -1 rather than -1.35, although the decay flux-distance relation is not as flat as it is for {\it XMM-Newton} and therefore the power law approximation is worse. For this instrument, the scales probed are typically within the region where the density profile slope is shallower than -2, rather than steeper as was the case for {\it XMM-Newton}, thus the extra dark matter enclosed within the FoV is larger with increasing distance and partially offsets the decrease in flux. We have considered the case of the {\it ATHENA/XIFU} FoV, which is intermediate in size between the previous FoV, and find the best power law approximation index is -1.1. 
 
 Finally, we considered the case of fixed physical apertures -- 8~kpc, 16~kpc and 30~kpc -- as opposed to the fixed opening angle above for Ref-L100N1504 and Rec-L25N752. We find that the flux from an 8~kpc aperture drops off with a power law index of -1.9,  and at 30~kpc the index is -2.0, and thus the same as a point mass.  

  \subsection{Local group analogue systems}
  \label{subsec:m31}

In this Section we consider observations of three constituent galaxies/galaxy classes of the Local Group \citep{Fattahi15}: the flux profile of M31, dwarf galaxies at the distance of M31 (including, but not limited to, M31 satellites) and MW satellites. In the final two cases we also consider the effect of the dark matter model, CDM versus WDM.

\subsubsection{M31 flux profile}

The M31 galaxy is of particular interest to X-ray decay studies due to its extent on the sky: we can take pointings at multiple radii to examine whether the measured signal is well described by a dark matter profile as would be the case for a dark matter decay line, or instead by a profile that traces the gas and thus disfavours a dark matter interpretation. The small scales probed by these observations in such a nearby object, of the order of parsecs, imply that measurements are sensitive to the effect of baryons on the dark matter halo as illustrated in Fig.~\ref{MvXDMO}.  

\begin{figure}
  	\includegraphics[scale=0.55]{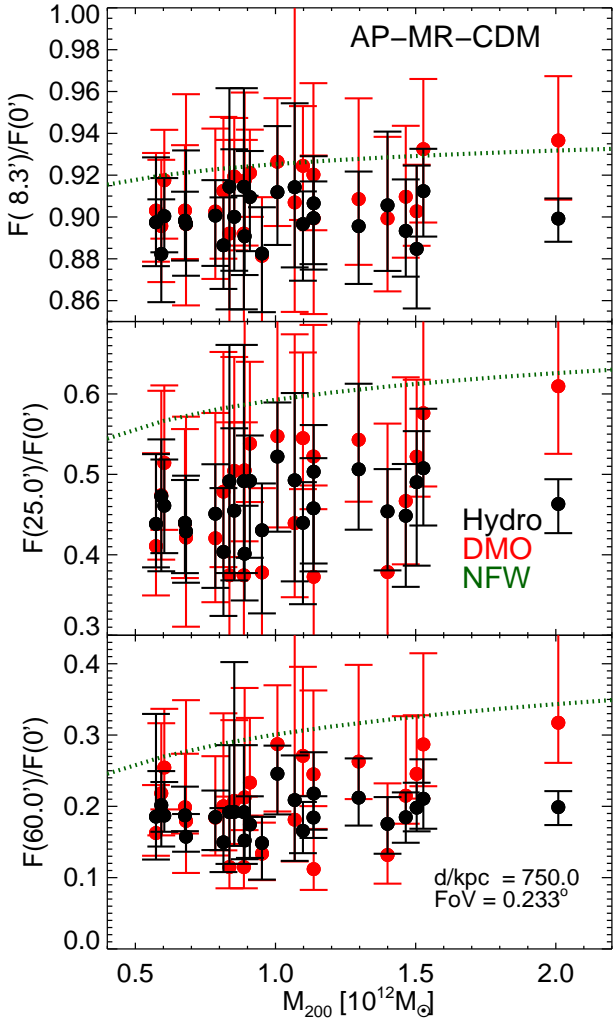}
    \caption{Ratio of decay flux relative to the flux on-centre with offset for M31 candidate haloes at the distance of M31 as a function of halo mass. The three offset angles are 8.3' (top panel), 25.0' (middle panel), and 60.0 (bottom panel). The points mark the medians of the flux ratios for each observer and the error bars denote the 95~per~cent data range. Data from the hydrodynamical simulations are shown in black, and from the DMO counterparts in red. The semi-analytic NFW flux ratio is shown as a green dotted line.}
    \label{M31MhvO}
  \end{figure}

We consider four pointings, at displacements from the centre of M31 of 0.0', 8.3', 25.0' and 60.0' made for a distance to M31 of 750~kpc \mbox{\citep{McConnachie05}}. We select the two largest simulation haloes in each AP-MR (CDM) simulation to be our M31 analogues for a total of 24 M31 analogues \footnote{The APOSTLE volumes are chosen to host a pair of galaxies that have the approximate halo mass of M31 and the MW, and with the same separation as the measured M31-MW distance. We treat both the M31 and MW-analogues as M31-like galaxies.}. We generate 500 observers placed randomly on the surface of a spherical shell of radius 750~kpc around the M31-analogue centre, and for each of those perform the four virtual pointings. We then compute the ratio of the three off-centre pointings to the on-centre observation, compute the median and 95~per~cent range across the 500 virtual observations, and plot the results as a function of halo virial mass in Fig.~\ref{M31MhvO}. We also include results for the same set of observers and pointings when using the DMO versions of the APOSTLE simulations, plus the NFW profile that assumes the Ref-L100N1504 dark halo concentration-mass relation (pink dotted line).

The suppression of each off-centre flux relative to the flux at the centre is approximately 0.9, 0.45, and 0.2 for 8.3', 25.0', and 60.0' respectively. There is a weak trend for the degree of suppression to decrease as a function of increasing halo mass, due to the anti-correlation of concentration with halo mass, but this trend is subdominant to the uncertainty induced by different viewing angles of the same halo, which is of the order of a few per~cent at 8.3', tens of per~cent at 25' and a factor of two at 1$^\rmn{o}$. Also remarkable is the effect of the baryons on the average suppression, which contributes a few extra per~cent in all three panels due to contraction of the halo compared to the DMO halo data (red points). Even when we assume the hydrodynamical EAGLE-derived NFW profile we underestimate the suppression by up to 10~per~cent, thus reflecting the limitations of the NFW profile in describing the matter distribution inside EAGLE galaxies as found by \citet[][fig.~10]{Schaller15}. Finally, we note that we have repeated this exercise with stellar mass instead of halo mass, and find that there is no clear trend in the decay flux ratio with stellar mass. We conclude that predictions for the M31 radial flux profile are sensitive to baryon physics, and are steeper than predicted by the NFW profile.

  \subsubsection{M31 satellites: effect of warm dark matter}

Dark matter models in which the dark matter undergoes decay typically belong to the WDM class of models. Low mass haloes ($<10^{11}\Msun$) in which the dark matter is warm have lower central ($<2$~kpc) densities than in CDM \mbox{\citep{Lovell14,Bose16a}}, and so the expected decay signal will be suppressed. Therefore, we perform virtual observations of WDM simulations as well as CDM in order to measure the extent of this suppression due to WDM.

The halo mass-concentration relation will vary as a function of the precise WDM properties. The primary model of interest to us -- due to its potential as an origin for the 3.55~keV line \citep{Boyarsky14a,Boyarsky15,Bulbul14,Cappelluti17} and ability to match Local Group galaxy properties \citep{Bozek16,Lovell17b,Lovell17a} -- is the decay of a 7~keV resonantly produced sterile neutrino. For the decay amplitude to be consistent with the measured fluxes at 3.55~keV for M31 and the GC, the mixing angle for this sterile neutrino must be in the range $[2,20]\times10^{-11}$, which corresponds to a lepton asymmetry, $L_6$ between 11.2 and 8 \citep{Laine08,Abazajian14,Boyarsky14a,Lovell16}. In order to maximise the likely flux suppression due to a 7~keV sterile neutrino candidate, we use simulations in which $L_6=11.2$ as this is the model with the largest free-streaming length\footnote{A 7~keV thermal relic particle could also decay and produce this signal. Its free-streaming length is much smaller than that of any 7~keV sterile neutrino, and thus the X-ray decay flux distribution would be indistinguishable from a decaying CDM particle.}. 

\begin{figure}
  	\includegraphics[scale=0.33]{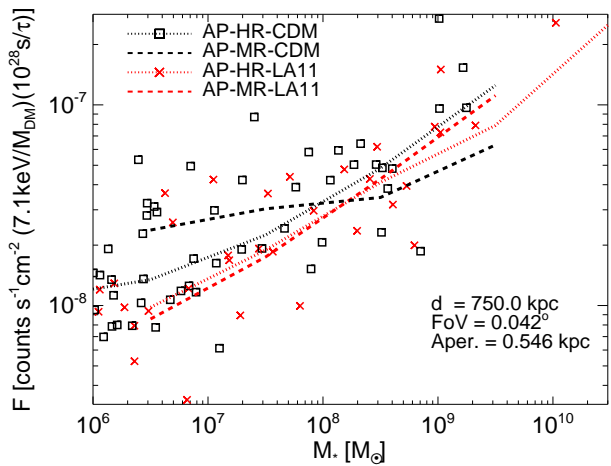}
    \caption{M31 satellite decay flux as a function of stellar mass for CDM (black) and the 7~keV sterile neutrino (red), at an observer distance of 750~kpc. Individual galaxies in the AP-HR-CDM and AP-HR-LA11 simulations are shown as squares (CDM) and crosses (LA11). The median decay flux-stellar mass relations of the high-resolution and medium-resolution simulations are shown as dotted and dashed lines respectively.}
    \label{M31MsWDM}
  \end{figure}

We measure the extent of the flux suppression in the context of our Local Group observations using one of the APOSTLE volumes simulated with both CDM and the 7~keV/$L_6=11.2$ sterile neutrino. We select all available galaxies in the simulation, both satellites and isolated galaxies, that have at least 100 star particles and 100 bound dark matter particles, and perform 500 virtual observations at a distance of 750~kpc. Many of these galaxies have dark matter masses as low as $10^{9}\Msun$ and are thus susceptible to numerical noise ($\sim10^{4}$ particles for the medium-resolution simulations). We therefore consider the medium- (MR) and high-resolution (HR) versions of each simulation in order to test for differences with resolution; we also adopt the {\it ATHENA/XIFU} FoV, which gives us an aperture radius at the target galaxy distance of $\approx1.1$~kpc. We present the median flux  -- out of the 500 observations -- as a function of stellar mass for this galaxy sample in Fig.~\ref{M31MsWDM}. For the high-resolution simulation data we plot both the flux for individual galaxies and the median flux-stellar mass relation, whereas for the medium-resolution counterparts we only plot the median relation.

There is scatter in the high resolution data of $\log{F/F_\rmn{3.55keV}}=\pm0.4$ at $10^{8}\Msun$, and the amplitude of the scatter grows towards lower masses. The median relation for the high-resolution WDM simulation is suppressed by $\sim10$~per~cent relative to CDM, although this is much smaller than the scatter of the points and therefore requires further statistics to be confirmed as significant. The medium resolution simulation is in reasonable agreement with its high resolution counterpart for $M_{*}<10^{9}\Msun$, whereas in the CDM case medium resolution returns a shallower relation than high resolution, suggesting that again small number statistics is affecting our results. Part of the reason for the agreement between resolutions despite the small aperture size is that we include the decay flux contribution from dark matter between the observer and the satellite, which we discuss further in the MW satellite context. We conclude that the nature of the dark matter has a minor impact on the fluxes measured for M31 satellites. 

 \subsubsection{MW satellites: effect of warm dark matter}

\begin{figure}
  	\includegraphics[scale=0.33]{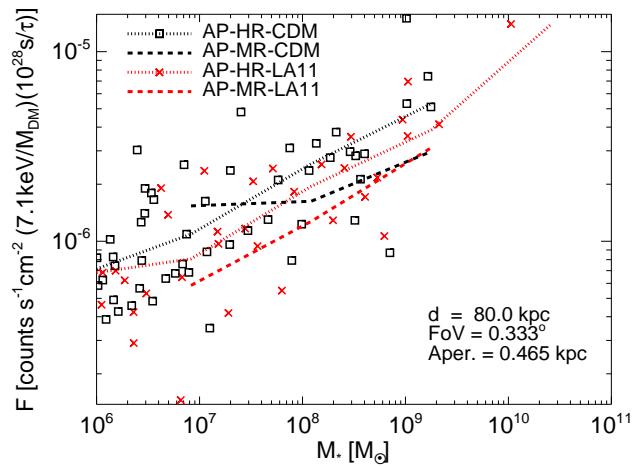}
    \caption{MW satellite decay flux as a function of stellar mass for CDM (black) and the 7~keV sterile neutrino (red), at an observer distance of 80~kpc. Individual galaxies in the HR simulations are shown as squares (CDM) and crosses (LA11). The medians of the high and intermediate data points are shown as dotted and dashed lines respectively.}
    \label{MWMsWDM}
  \end{figure}

A more challenging class of targets, from the point of view of virtual observations of simulations, is the Milky Way satellite population. Their close proximity to an observer on Earth -- typically 50-100~kpc and thus on average ten times closer than the M31 satellites -- means that even large FoV probe a small region of the halo centre, where the effects of limited resolution ($\lsim1$~kpc), dark matter physics \citep[$\lsim3$~kpc,][]{Lovell14}, and baryonic feedback are expected to be more prominent. We therefore repeat the exercise shown in Fig.~\ref{M31MsWDM} for MW satellites. We select our target galaxies to be isolated and satellite galaxies that have at least 100 star particles and 100 bound dark matter particles. We place our galaxies at 80~kpc from the observer with the {\it ATHENA/WFI} FoV, for an aperture at the target of 470~pc; we note that \citet{Neronov16} have shown that {\it ATHENA/XIFU} is also an excellent instrument for detecting the line in MW dwarf spheroidals, but our simulation resolution is insufficient at the {\it ATHENA/XIFU} FoV. We generate 500 virtual observations, and select the lowest flux of the 500 measured in order to reduce as far as possible the contribution of the MW main halo; there is therefore one data point per target galaxy. To simulate a complete observational signal it will be necessary to add on a MW halo component separately, which we leave to future work: here we are interested instead in studying the difference between WDM and CDM within the dwarf galaxies independent of their location with the MW halo. The results are presented in Fig.~\ref{MWMsWDM}.

There is an apparent shift in the median decay flux in the sterile neutrino model compared to CDM, of around 30~per~cent for galaxies with $M_{*}<10^{8}\Msun$ between red and black dotted lines. This difference is similar at lower resolution, although the statistical power in this small dataset, especially in the context of systematics associated with the baryon physics model, is insufficient to say definitively that the two distributions are different. Also, we note that there is a systematic offset between the two resolutions of the LA11 satellites, showing that resolution has not been achieved and so our results should be treated as a lower limit. Unlike the M31 satellites, there is no large mass of intervening dark matter in each sightline to compensate for the poor resolution; we note that adding a MW halo component will make the WDM-CDM difference smaller still. 

We therefore anticipate that further work with more simulations will make a key prediction specifically for sterile neutrino dark matter as a source of the 3.55~keV line: that the fluxes measured for MW satellites are suppressed by up to 30~per~cent relative to what one would have expected from an extrapolation of the decay flux--stellar mass relation calibrated for distant, massive galaxies.

   \subsection{The Perseus cluster}
   \label{subsec:perseus}

Another target of interest is the Perseus galaxy cluster. This target has the appeal of being a large dark matter mass that is relatively nearby ($\sim70$~Mpc) and can hence be probed as a function of radius. In this Section we examine the flux profiles and FWHM measurements of Perseus-analogues drawn from the C-EAGLE simulations, where our definition of a Perseus-analogue cluster is simply a halo with $M_{200}>10^{14}\Msun$ placed at a distance of 69.5~Mpc. The value of $M_{200}$ for Perseus inferred from X-ray spectroscopy by \mbox{\citet{Simionescu11}} is $6.65^{+0.43}_{-0.46}\times10^{14}\Msun$, and we make reference to this estimate in our plots. We use the {\it XMM-Newton} (Figs.~\ref{PerseushvO},\ref{PerseussvO}) and {\it XRISM} (Fig.~\ref{PerseussvO}) FoV to measure the flux as a function of radius, and then apply the {\it XRISM} FoV also to measure the FWHM given the anticipated excellent spectral resolution of that instrument ($<600$~\kms, Fig.~\ref{PerseusFWHM}).

\subsubsection{Surface brightness profiles}

\begin{figure}
  	\includegraphics[scale=0.55]{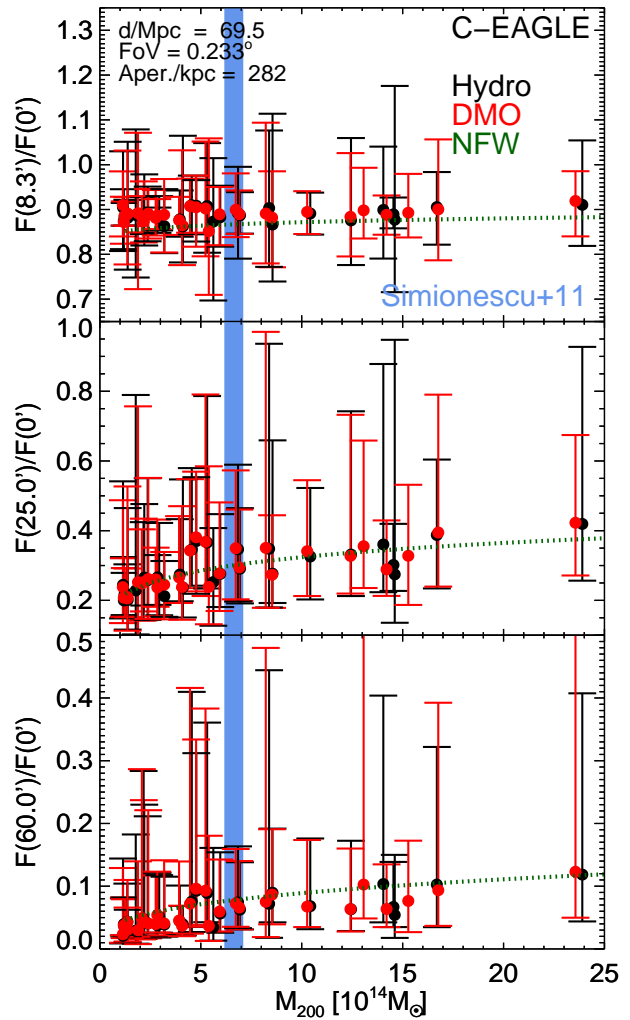}
    \caption{Ratio of flux compared to central flux at various offsets from the centre of simulated Perseus analogues at the Perseus distance as a function of halo mass. The three offset angles are 8.3' (top panel), 25.0' (middle panel), and 60.0' (bottom panel). We show data from the hydrodynamical runs in black and from the DMO counterparts in red. The points show the median of each distribution of flux and the error bars the 95~per~cent range. The $1\sigma$ uncertainty on the mass of Perseus as measured by \mbox{\citet{Simionescu11}} is shown as a vertical blue band. The NFW semi-analytic relations using the Ref-L100N1504 mass-concentration relation are shown as dotted green lines.}
    \label{PerseushvO}
  \end{figure}

We repeat the process that we applied to our M31 haloes in Fig.~\ref{M31MhvO} but now use the C-EAGLE haloes, which we place at a distance of 69.5~Mpc.  Our three offset angles are 8.3', 25.0', and 60.0' (which are 9, 27 and 66 per~cent of the Perseus $r_{200}$ at the Perseus distance). We plot the range of flux ratios from each virtual observation as a function of $M_{200}$ in Fig.~\ref{PerseushvO}. 

  The average suppression relative to the flux at the centre as a function of offset angle is 0.90, 0.3, and 0.03 for angles of 8.3', 25.0', and 60.0' respectively. The variation between different viewing angles is large, with some 8.3' offset observations returning a higher flux than the on-centre measurement, possibly due to substructure. For all three offset angles there is a tendency towards higher ratios at higher masses, 0.35 at $1.5\times10^{12}\Msun$ compared to 0.25 for our lowest-mass haloes at 25.0'. The proportion of relaxed haloes decreases as halo mass increases \citep{Neto07} so we expect the variation between sightlines of the same object to be greater in clusters. In the same figure we include results when observing the same volumes, with the same sightlines, of the DMO counterpart simulations. We do not see any systematic trend from the hydrodynamic simulations to differ from either the DMO simulations or the NFW result, which is due to the large aperture subtended by the FoV at this distance ($\sim280$~kpc radius) averaging over the regions in which halo contraction occurs.
  
\begin{figure}
  	\includegraphics[scale=0.55]{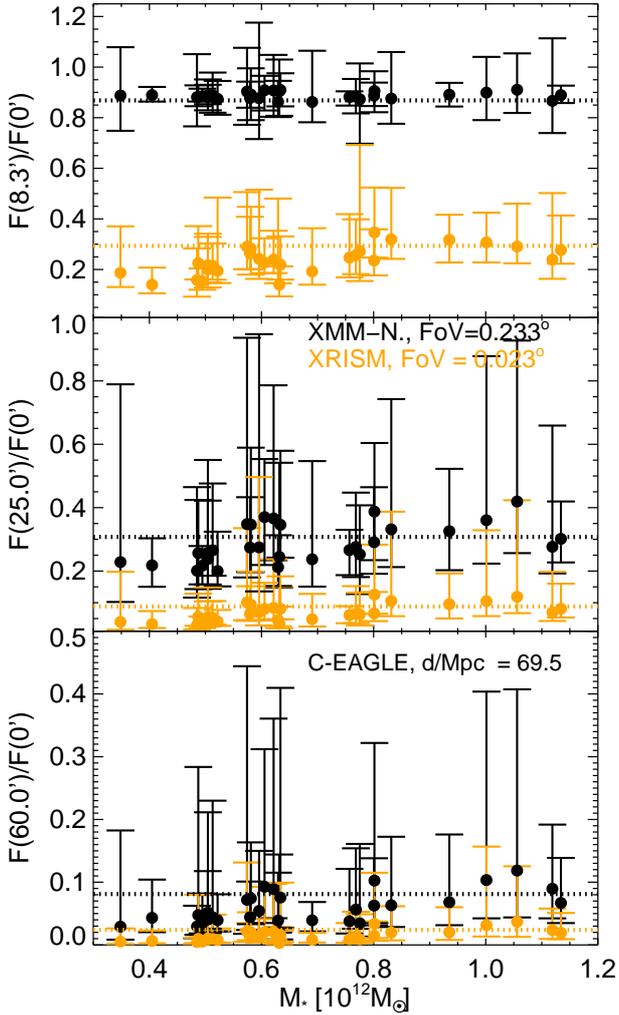}
    \caption{Ratio of flux compared to the central flux at various offsets from the Perseus candidate haloes at the Perseus distance as a function of stellar mass. The three offset angles are 8.3' (top panel), 25.0' (middle panel), and 60.0' (bottom panel). Predictions for the {\it XMM-Newton} FoV are shown in black and for {\it XRISM} in orange. Points mark the median of the data and the error bars denote the 95~per~cent range. The dotted lines show the flux ratios for an NFW halo of $7\times10^{14}\Msun$ -- the mass of Perseus as measured by \citet{Simionescu11} -- for {\it XMM-Newton} and {\it XRISM} in their corresponding colours. Note that the {\it y}-axis ranges are different for each panel.}
    \label{PerseussvO}
  \end{figure}

For the hydrodynamical runs we can repeat the analysis of flux offset as a function of stellar, rather than halo, mass (Fig.~\ref{PerseussvO}). We also consider similar observations for the {\it XRISM} FoV, which is smaller than its {\it XMM-Newton} counterpart and therefore probes the flux profile in greater detail (28~kpc radius). We further plot the values of the ratios of the Perseus mass ($\sim7\times10^{14}\Msun$) NFW profile for both FoV as dotted lines. There is a similar trend of the 25.0' and 60.0' flux ratios to increase with stellar mass, but again the asphericity of the halo and its environment dominates, as reflected in the scatter of individual haloes.  

  Based on all the results of this Subsection, we conclude that the greatest uncertainty on the radial profile is the asphericity of Perseus ($\sim10$~per~cent) rather than the effects due to galaxy formation, the halo mass, or the stellar mass-halo mass variation. The {\it XRISM} virtual observations show a much greater decline with radius than is the case for the {\it XMM-Newton} FoV: a suppression of 0.2 at 8.3', 0.05 at 25.0' and $<0.05$ at 60.0'. This is due in part to the smaller FoV not picking up flux from the inner parts of the halo in the offset measurement, and also perhaps due to contraction of the dark matter halo within the central galaxy ($<30$~kpc) as discussed below in the context of the FWHM. We show that the {\it XRISM} flux ratios are lower than the NFW profile whereas the {\it XMM-Newton} flux ratios are not, and have checked that the 8.3' to 0' flux ratio for the DMO C-EAGLE haloes is of the order of 10~per~cent higher than for their hydrodynamical counterparts (not shown). We caution that the degree of contraction in C-EAGLE may be stronger than any that occurs in the real Universe, as the C-EAGLE BCGs are 2-3 times more massive than their observed counterparts \citep{Bahe17}.
  
\begin{figure}
  	\includegraphics[scale=0.55]{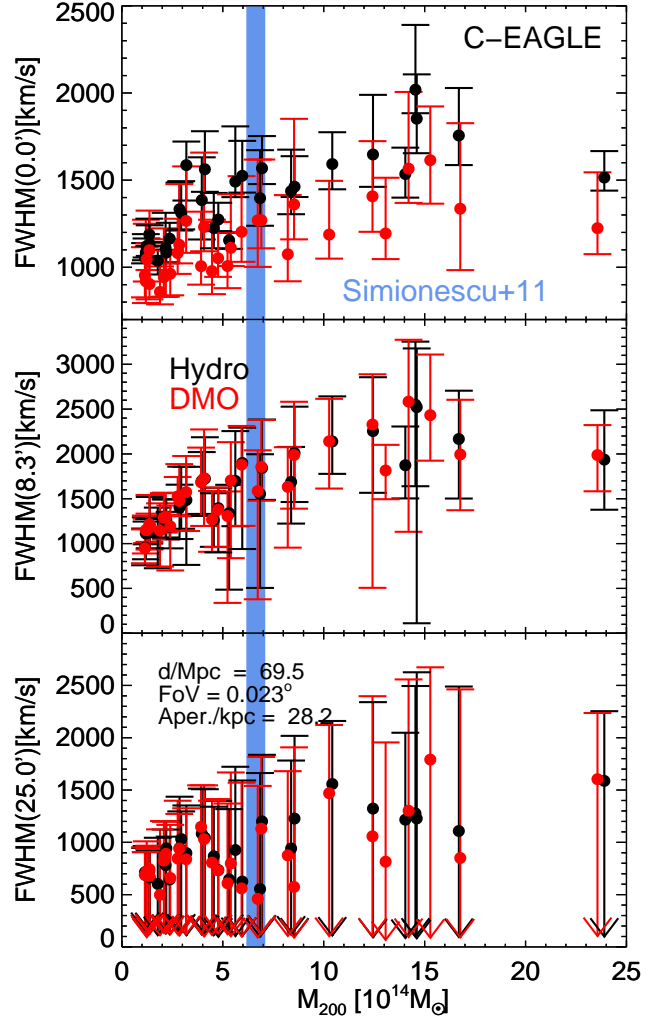}
    \caption{The FWHM of the flux measured for different sightlines in our Perseus virtual observations as a function of halo mass while using the {\it XRISM} FoV. We display results for on-centre observations (top panel) and at offsets of 8.3' (middle panel) and 25.0' (bottom panel). Data from the hydrodynamical simulations are shown in black, and those from the DMO simulations in red. The error bars enclose the 68~per~cent range. The $1\sigma$ uncertainty on the mass of Perseus as measured by \citet{Simionescu11} is shown as a vertical blue band. In the bottom panel the lower bound of the 68~per~cent range for each halo is no higher than the minimum FWHM that we resolve, 140~\kms, therefore we mark these lower bounds with arrows rather than an error bar hat. }
    \label{PerseusFWHM}
  \end{figure}

  We conclude our study of Perseus with an analysis of the expected velocity width of the dark matter decay line. The width of the line is  determined by the velocity dispersion of the host halo within the FoV, which is higher than that of the hot gas in the central regions of clusters that also emit lines since dark matter has no cooling mechanism. A broad line is thus a signature of dark matter. We measure the line width within three of our offsets (0.0', 8.3' and 25.0') for the {\it XRISM} FoV. For each of the particles enclosed in the FoV we calculate the velocity component along the line of sight and bin up the flux from all particles in bins of width $\sim$70~\kms. We compute the FWHM of the resulting velocity distribution and, in turn, obtain a distribution of FWHM across the 500 sightlines for each halo. We plot the median and 68~per~cent range of these data in Fig.~\ref{PerseusFWHM}, for both the hydrodynamical and DMO versions of each halo.      
  
 The measured FWHM increases with halo mass from $\approx1100$~\kms at $1\times10^{14}\Msun$ to $\sim2000$~\kms at $1.5\times10^{15}\Msun$ in the hydrodynamical simulation for the on-centre observations. In the $M_{200}$ range measured by \citet{Simionescu11}, the measured FWHM lies in the range [1300,1700]~\kms, which is a factor of two larger than obtained from a similar calculation performed for the gas particles ([150,550]~\kms taking into account bulk and thermal velocities; not shown). A larger increase with $M_{200}$ occurs for the 8.3' offset observations, up to $\sim2500$~\kms at $1.5\times10^{15}\Msun$. The 25.0' offsets show much larger variations between sightlines because of lower dark matter flux; the FWHM clearly increases with halo mass. The most conspicuous difference between the on-centre and two off-centre observations is the enhancement of the FWHM due to baryonic physics, by up to 50~per~cent in some cases for the on-centre observations but nearly zero for the off-centre observations; we expect this result is due to contraction of the halo discussed above. Finally, the variation in the 68~per~cent range is generally of the order of tens of per~cent but occasionally much larger; the 95~per~cent ranges (not shown) encompass factors of two or more.  
 
 We note that the {\it Hitomi} collaboration used 1300~\kms as the fiducial upper limit for the FWHM of the dark matter decay line, which as we see may be underestimated by 50~per~cent or more. Even for this line width the dark matter signal interpretation of 3.55~keV signal was consistent with the {\it Hitomi} non-observation at about $3\sigma$ (energy dependent, see \citealp{HitomiC16}). For wider lines the non-detection becomes fully consistent. This should be contrasted with Hitomi limit on the presence of a \emph{Potassium} atomic line in this range, as the latter is expected to have an order of magnitude narrower FWHM. Therefore the {\it Hitomi} observation \emph{rules out} interpretation of the 3.55~keV signal from Perseus cluster as an atomic line \citep{Jeltema15} but does not contradict the dark matter interpretation. 
 
 In conclusion, we have measured the flux profiles of candidate Perseus haloes. We have mapped the suppression of the X-ray decay flux as a function of observation offset angle, and have shown that this suppression correlates weakly with both halo mass and stellar mass; the decline is steeper for the {\it XRISM} FoV. We have also predicted the FWHM of the line measured with {\it XRISM}  and found that the on-centre FWHM measurement is enhanced by tens of per~cent by the influence of baryon physics.

	\subsection{Distant clusters}
    \label{subsec:hrc}

One of the first studies to report a possible detection of the previously unknown 3.55~keV line was based on stacked clusters \citep{Bulbul14}, with redshifts in the range $z=[0.009,0.35]$. This approach has the benefit of smearing out instrumental lines, which will shift in velocity relative to the redshifted line in the target, and in principle leave behind only those lines associated with the target cluster. In this subsection we use our C-EAGLE halo set to construct a sample of haloes distributed across cosmic time, taking advantage of the different snapshot outputs to examine the same haloes at various stages of their evolution. We choose three redshifts: $z=0.016$ -- the redshift of the Perseus cluster, -- $z=0.1$ and $z=0.25$. These latter two redshifts are the two available simulation outputs below that of the most distant cluster in the cluster sample of \citet[][$z=0.35$]{Bulbul14}. 

\begin{figure}
  	\includegraphics[scale=0.35]{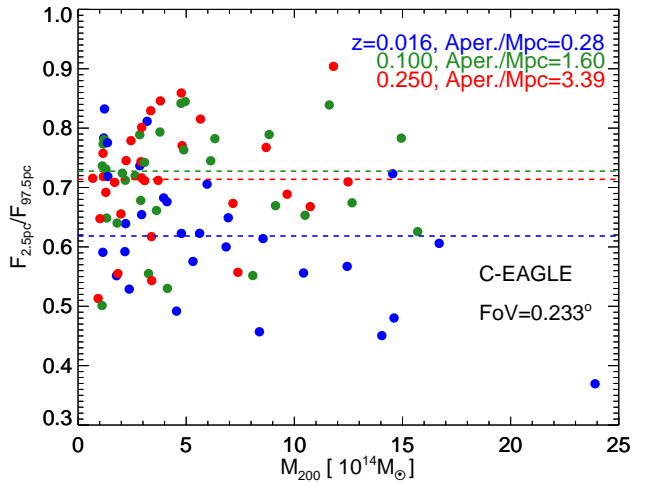}
    \caption{Ratio of minimum to maximum fluxes measured for C-EAGLE cluster haloes as a function of $M_{200}$ measured at redshifts 0.016, 0.10 and 0.25 (blue, green and red symbols respectively).}
    \label{HRC}
  \end{figure}

Two properties of interest for clusters are the flux amplitude and the scatter due to the viewing angle. We present the expected flux in the following Section; here we restrict our attention to the scatter. We compute 500 orthogonal sightlines for each halo and then calculate the ratio of the fluxes that enclose 95~per~cent of the data (the 2.5~per~cent and 97.5~per~cent highest fluxes.) For this part of the analysis we modify the spherical aperture for including particles around the target, since the 2~Mpc aperture fits entirely within the {\it XMM-Newton} FoV at z=0.25 (3.4~Mpc). We therefore increase the size of the aperture to 10~Mpc (proper distance). We present the results in Fig.~\ref{HRC}.

Almost all of the $z=0.016$ targets show less than 50 per~cent variation in flux between viewing angles, and there is potentially a trend for more massive haloes to show a bigger variation as one would expect if they are less relaxed; the median suppression is 40~per~cent. We have checked this result against a repeat calculation using the previous 2~Mpc spherical aperture and find that the difference between the 2~Mpc and 10~Mpc fluxes at $z=0.016$ is negligible. The $z=0.1$ and $z=0.25$ haloes show a smaller variation than the $z=0.016$, both around 30~per~cent median suppression, and we expect that this is due to the larger size of the FoV relative to the cluster virial radius.

  \section{Conclusions}
  \label{sec:conc}
  
    \begin{figure*}
  	\includegraphics[scale=0.7]{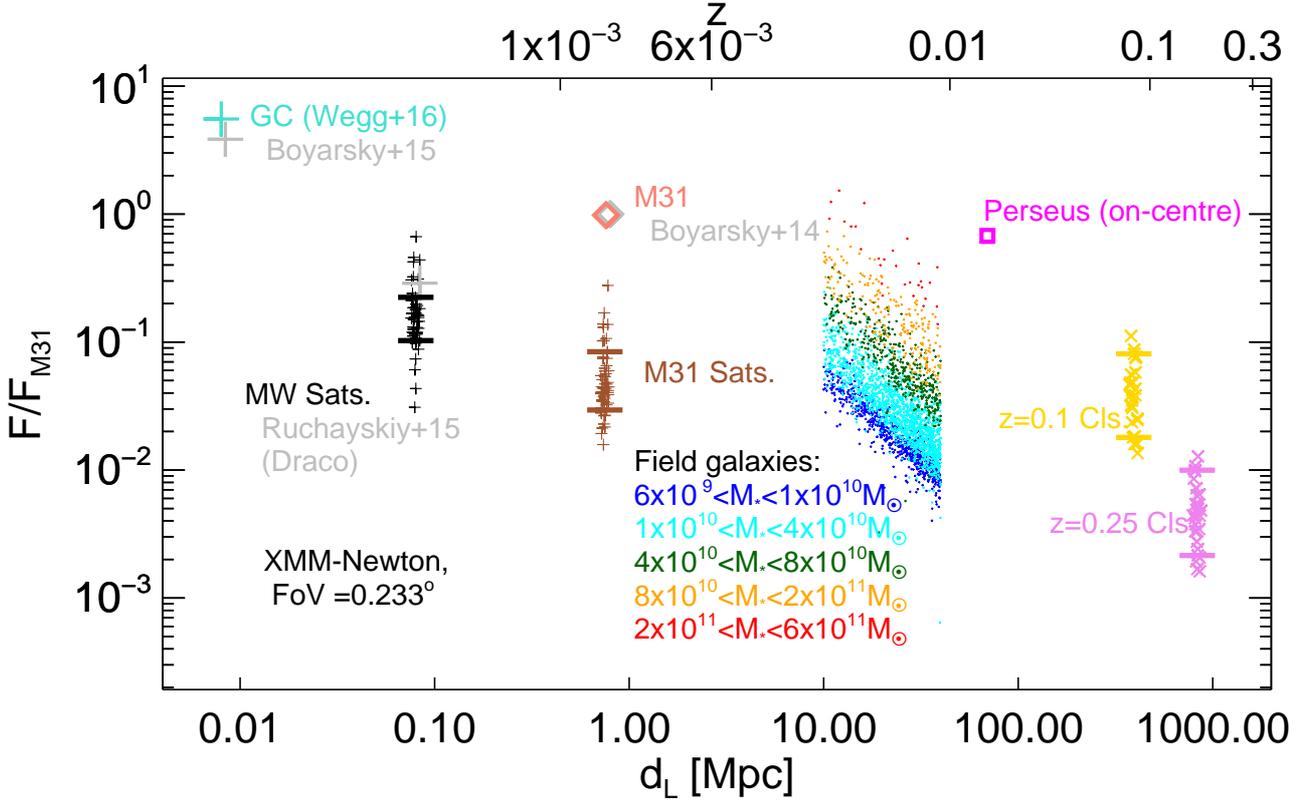}
    \caption{The predicted flux for various targets as a function of distance to the target, where the flux is measured using the {\it XMM-Newton} FoV and then normalised by the measured 3.55~keV line flux from M31 \citep{Boyarsky14a} . We show MW satellites as black pluses (80~kpc), M31 satellites as brown pluses (750~kpc), the $z=0.1$ clusters as gold crosses and the $z=0.25$ clusters as violet crosses. The distance to each target in these sets is multiplied by a random number of up to 10~per~cent for clarity. M31 itself is shown as a pink diamond, and the centre of Perseus as a magenta square. The field galaxies are shown as dots and are separated into five bins in stellar mass by colour as indicated in the plot legend. The fluxes are computed at 20~Mpc, assigned a new position uniformly distributed between 10 and 40~Mpc and then multiplied by the relation in Fig.~\ref{MvXDistR} to obtain the expected flux at the new redshift. For the MW satellites, M31 satellites, $z=0.1$ clusters and $z=0.25$ clusters the 68~per~cent region of the data is delineated by two horizontal lines.  The M31 satellites and MW satellites are drawn from the WDM high resolution APOSTLE, and M31 itself is the most massive galaxy in WDM high resolution APOSTLE. The cluster samples at $z=0.1$ and $z=0.25$ are all of the clusters in C-EAGLE, and Perseus is one of the two C-EAGLE clusters that agrees with the measured Perseus $M_{200}$. The field galaxies are drawn from EAGLE L100-N1504. We also include an estimate of the GC decay signal as derived from microlensing observations by \citet{Wegg16}: note that this point does not use any simulation data. Finally, we add three observational data points in grey: the claimed GC and M31 detections presented in \citet{Boyarsky15} and \citet{Boyarsky14a} respectively, and the reported $2\sigma$ excess in the Draco dSph \citep{Ruchayskiy15}.}
    \label{SummaryPlotXMM}
  \end{figure*}
  
    \begin{figure}
  	\includegraphics[scale=0.35]{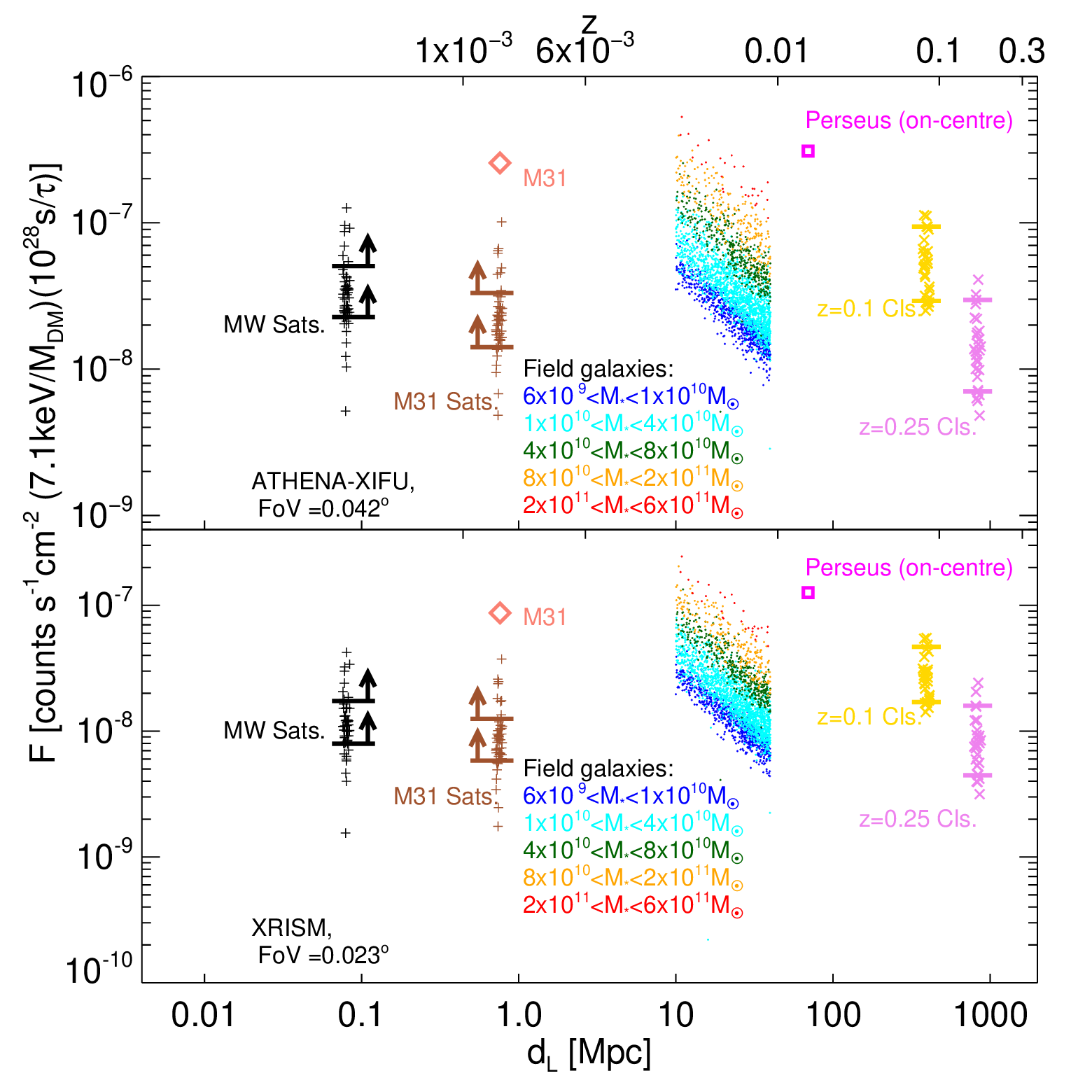}
    \caption{The predicted flux for each of the targets featured in Fig.~\ref{SummaryPlotXMM}, but using the {\it ATHENA/XIFU} (top panel) and {\it XRISM} (bottom panel) FoV. Note that the MW satellite and, to a lesser degree, M31 satellite measurements suffer from poor resolution at these small scales and thus represent a lower limit on the expected flux, which we represent with arrows in the Figure. The power law used for the drop off field galaxies is -1 as derived for {\it XRISM} in Section~\ref{sss:dist}. }
    \label{SummaryPlotBoth}
  \end{figure}

  We have used simulations of galaxy formation to make predictions for the signal from decaying dark matter. We have taken advantage of the broad scope of the EAGLE project and its daughter projects, APOSTLE and C-EAGLE, to measure the likely amplitude, scatter, and in some cases full width-half maximum (FWHM), of the decay flux line from a series of objects that differ by four orders of magnitude in distance scale, to six orders of magnitude in stellar mass, and six orders of magnitude in dark matter halo mass; from Milky Way satellite galaxies to massive clusters at redshifts up to $z=0.25$. 
  
  In this way we have generated a series of constraints, which should be useful to assess the validity of a detection of dark matter decay. In particular, we show that the FWHM of the line originating from the Perseus-sized cluster is on average in the range 1300-1700~\kms (and can exceed 2000~\kms in some realisations).
Therefore, the non-detection of the 3.55 keV line by the {\it Hitomi collaboration} is consistent with its DM interpretation -- the collaboration used 1300~\kms as a fiducial upper bound on the line width. At the end of this section we summarize our results with a comparison to  the 3.55~keV line amplitude  measured from existing observations. We also predict signals for future X-ray missions such as {\it XRISM} and {\it ATHENA}, and identify relations between potential signals coming from different (types of) objects.  
  
We began with an analysis of galaxies observed at a fixed `fiducial' distance of 20~Mpc, with a focus on field galaxies (Fig.~\ref{MvX}). We performed three virtual observations of $>11,000$ simulated galaxies in the stellar mass range [$10^{7},10^{12}]~\Msun$ across all of the simulations. We showed that the $1\sigma$ halo-to-halo scatter around the median flux is approximately 30~per~cent (Fig.~\ref{MvXR}). The $1\sigma$ variation due to viewing angle is 20~per~cent for bright ($M_{*}>10^{9}\Msun$) galaxies, but can be as much as 60~per~cent at the 95~per~cent contour. The variation is stronger for less massive galaxies, 35~per~cent at the 68~per~cent contour, which can indicate both a more aspherical halo and the intrusion of other, relatively massive haloes into the field-of-view (FoV). However, the consistently largest source of systematic uncertainty is related to the baryon physics included in the model, where different choices of how to calibrate the baryon physics model affect the stellar mass-halo mass relation and change the expected median flux by 40~per~cent at fixed stellar mass (c.f. AP-MR versus the recalibrated Rec-L25N752, Fig.~\ref{MvXS}).

A further source of uncertainty is the impact of the baryons on the dark matter distribution within galaxies. We found that galaxies with $M_{*}>10^9\Msun$ were progressively more concentrated when baryons were included, i.e. compared to their dark matter-only (DMO) simulation counterparts, and therefore the measured decay flux was enhanced up to 40~per~cent at $M_{*}=2\times10^{10}\Msun$ within a 4~kpc aperture (Fig.~\ref{MvXDMO}). We considered the role of environment, and found that nearby haloes contribute to the flux measured within an aperture of 80~kpc about the target galaxy centre by up to 40~per~cent for galaxies with $M_{*}<10^{10}\Msun$; however, measurements within apertures of 8~kpc are not affected by the local environment (Fig.~\ref{MvXFR}).    

We considered sources of scatter in X-ray decay flux at fixed stellar mass. We showed that halo mass is a strong source of scatter for galaxies located at 20~Mpc from the MW (81~kpc aperture), and that this scatter is mirrored by the abundance of bright satellites (Fig.~\ref{Sep4Panel}). The halo concentration plays a more complicated role, with more concentrated haloes showing greater fluxes than their less concentrated counterparts for $M_{*}<10^{10}\Msun$ in the central regions of galaxies (8~kpc aperture) but the opposite is true for the full 81~kpc apertures. We also showed that galaxies with older stellar populations presented larger decay fluxes in the central 8~kpc. 

We concluded our discussion of field galaxies by making predictions for the flux as a function of galaxy distance. First, we showed that at fixed stellar mass the flux within the FoV of {\it XMM-Newton} falls off between 10~Mpc and 40~Mpc with an approximate power law of $-1.35$ in the mass range $10^{7}<M_{*}<10^{10}\Msun$ (Fig.~\ref{MvXDistR}), which is very similar to that predicted by the Navarro--Frenk--White profile \mbox{\citep[NFW,][]{NFW_96,NFW_97}}; the fall off is shallower for extreme masses either side. Finally, we showed that the flux of the halo in the region where the -1.35 power law applies is well approximated by the expression $d^{-1.35}x^{\gamma}(1+x)^{1-\gamma}$ for the {\it XMM-Newton} FoV, where $\gamma=0.3$ and $x=M_{*}/2\times10^{10}\Msun$.

Our second set of galaxies were those in the Local Group: M31, the satellites of M31 and the satellites of the Milky Way (MW). We showed that the close proximity of M31 enables us to detect the contraction of the halo due to baryons using {\it XMM-Newton}, such that the flux profile is steeper than inferred from DMO simulations (Fig.~\ref{M31MhvO}). Also, the variation with viewing angle is consistently larger than the variation in the profile with either halo mass or stellar mass. We then considered satellite galaxies of M31 and the MW, particularly in the context of warm dark matter (WDM), which are predicted to have lower central densities than their CDM counterparts. At the distance of M31, the size of the aperture subtended by the {\it ATHENA/XIFU} FoV is large enough that this density suppression is $\sim10$~per~cent (Fig.~\ref{M31MsWDM}), but dwarf galaxies observed at a the distance of MW satellites with the larger {\it ATHENA/WFI} FoV show  median fluxes of WDM satellites are suppressed up to the 30~per~cent level relative to CDM satellites (Fig.~\ref{MWMsWDM}), at least when the MW halo contribution to the decay flux is omitted.

We next considered Perseus galaxy cluster-analogue haloes. We showed that, with the FoV of {\it XMM-Newton}, baryons did not affect the flux profile, which, like that of M31, showed much greater scatter between sightlines than with halo mass (Fig.~\ref{PerseushvO}). The {\it XRISM} experiment will be able to measure the FWHM of any decay line. We showed that the expected FWHM to be measured in the centre of Perseus by {\it XRISM} is 1300-1700~\kms (68~per~cent), and is enhanced by $\sim20$~per~cent over the DMO expectation (Fig.~\ref{PerseusFWHM}). The measured FWHM at larger radii can be still higher, and is not affected by baryons.

The final set of objects that we considered is the general population of clusters, at redshifts of $z=0.016$ (Perseus), 0.1 and 0.25. We showed that the typical variation of flux between sightlines with 38~per~cent at the Perseus distance, 28~per~cent at $z=0.1$ and 29~per~cent for $z=0.25$ {Fig.~\ref{HRC}}. 

In summary, we have generated predictions for a population of galaxies and galaxy clusters at various stellar masses and distances, identifying the systematic shifts due to baryonic physics, uncertainty in the baryon model, and in stochastic variations between haloes. A crucial step is then to ascertain whether the signals measured for different objects, or ruled out to some confidence, are consistent with one another. We summarise all of our results in two plots, Figs.~\ref{SummaryPlotXMM}, and \ref{SummaryPlotBoth}. Here we show the predicted fluxes for all of the targets considered as a function of distance, from the MW satellites to the $z=0.25$ clusters. We also include a prediction for the Galactic Centre (GC, Fig.~\ref{SummaryPlotXMM} only) as inferred from the micro-lensing study of \citet{Wegg16} who found that the dark matter halo was well fit by an NFW profile with mass $1.1\times10^{12}\Msun$ and concentration $c=9$. We have based this prediction from observations because our simulations do not have the necessary mass resolution at these very small scales (see fig.~A2 of \citealp{Lovell15}).

The FoV used in the first case is that of {\it XMM-Newton}. In this plot we make some broad-stroke comparisons to the detections and upper limits reported by \mbox{\citet[][M31]{Boyarsky14a}}, \mbox{\citet[][Galactic Centre]{Boyarsky15}}, and \citet[][the Draco dSph, see \citealp{Jeltema16} for an alternative analysis]{Ruchayskiy15}, each of which is in good agreement with our results; we normalise the published measurements and detections by the 3.55~keV flux measurement of \citet{Boyarsky14a}. We repeat this exercise using the  {\it ATHENA/XIFU} and {\it XRISM} instruments, the latter of which has observational capabilities inferior to those of {\it ATHENA/XIFU} but is set to launch much sooner (2021 as opposed to 2028 at the earliest.) Future observations will map onto the various regions of these figures, and then provide decisive evidence of whether or not any unexplained X-ray line is indeed due to dark matter decay.

  \section*{Acknowledgements}

We would like to thank K\'ari Helgason, Oleg Ruchayskiy and Alexey Boyarsky for useful conversations.  This work was carried out on the Dutch National e-Infrastructure with the support of SURF Cooperative. MRL is  supported  by  a  COFUND/Durham  Junior Research Fellowship under EU grant 609412, and also acknowledges
support by a Grant of Excellence from the Icelandic Research Fund (grant number
173929$-$051). RAC is a Royal Society Research Fell ow. MS is supported by VENI grant 639.041.749. This work used the DiRAC Data Centric system at Durham University,
operated by the Institute for Computational Cosmology on behalf of the
STFC DiRAC HPC Facility (\url{www.dirac.ac.uk}). YMB acknowledges funding from the EU Horizon 2020 research and innovation programme under Marie Sk{\l}odowska-Curie grant agreement 747645 (ClusterGal) and the Netherlands Organisation for Scientific Research (NWO) through VENI grant 016.183.011. WAH is supported by an Individual Fellowship of the Marie Sk{\l}odowska-Curie Actions and therefore acknowledges that
this project has received funding from the European Union's
Horizon 2020 research and innovation program under the
Marie Sk{\l}odowska-Curie grant agreement No 748525. 

This equipment was funded by BIS National E-infrastructure capital grant ST/K00042X/1, STFC capital grant ST/H008519/1, and STFC DiRAC Operations grant ST/K003267/1 and Durham University. DiRAC is part of the National E-Infrastructure. The research was supported in part by the European Research Council under the European Union's Seventh Framework Programme (FP7/2007-2013)/ERC grant agreements, the National Science Foundation under grant no. NSF PHY11-25915, the UK Science and Technology Facilities Council (grant numbers ST/F001166/1 and ST/I000976/1), Rolling and Consolidating grants to the ICC. The C-EAGLE simulations
were in part performed on the German federal maximum performance computer `HazelHen' at the maximum performance computing centre Stuttgart (HLRS), under project GCS-HYDA / ID
44067 financed through the large-scale project `Hydrangea' of the Gauss Center for Supercomputing. Further simulations were performed at the Max Planck Computing and Data Facility (MPCDF) in Garching, Germany.
   
  \bibliographystyle{mnras}

\label{lastpage}

  \end{document}